\documentclass[%
 reprint,
superscriptaddress,
nobibnotes,
 amsmath,amssymb,
 aps,
floatfix,
]{revtex4-2}
\usepackage{graphicx}
\usepackage{dcolumn}
\usepackage{bm}
\usepackage{subfigure, tabularx}
\usepackage{float} 
\RequirePackage{afterpage}

\begin{document}


\preprint{APS/123-QED}

\title{The Role of Masks in Mitigating Viral Spread on Networks}

\author{Yurun Tian}
\affiliation{%
 Department of Electrical and Computer Engineering, Carnegie Mellon University, Pittsburgh, PA 15213 USA
}%
\author{Anirudh Sridhar}%
\affiliation{%
 Department of Electrical and Computer Engineering, Princeton University, Princeton, NJ, 08544 USA
}%


\author{Chai Wah Wu}
\affiliation{
 Thomas J. Watson Research Center,  IBM, Yorktown Heights, NY, 10598 USA
}%
\author{Simon A. Levin}
\affiliation{%
 Department of Ecology \& Evolutionary Biology,  Princeton University, Princeton, NJ, 08544 USA
}%
\author{Kathleen M. Carley}
\affiliation{%
 Software and Societal Systems, School of Computer Science, Carnegie Mellon University, Pittsburgh, PA 15213
}

\author{H. Vincent Poor}
\affiliation{%
 Department of Electrical and Computer Engineering, Princeton University, Princeton, NJ, 08544 USA
}%

\author{Osman Ya\u{g}an}
\affiliation{%
 Department of Electrical and Computer Engineering, Carnegie Mellon University, Pittsburgh, PA 15213 USA
}%


\date{\today}

\begin{abstract}
Masks have remained an important mitigation strategy in the fight against COVID-19 due to their ability to prevent the transmission of respiratory droplets between individuals. In this work, we provide a comprehensive quantitative analysis of the impact of mask-wearing. To this end, we propose a novel agent-based model of viral spread on networks where agents may either wear no mask or wear one of several types of masks with different properties (e.g., cloth or surgical). We derive analytical expressions for three key epidemiological quantities: the probability of emergence, the epidemic threshold, and the expected epidemic size. In particular, we show how the aforementioned quantities depend on the structure of the contact network, viral transmission dynamics, and the distribution of the different types of masks within the population. Through extensive simulations, we then investigate the impact of different allocations of masks within the population and trade-offs between the outward efficiency and  inward efficiency of the masks.
Interestingly, we find that masks with high outward efficiency and low inward efficiency  are most useful for controlling the spread in the early stages of an epidemic, while masks with high inward efficiency but low outward efficiency are most useful in reducing the size of an already large spread.
Lastly, we study whether degree-based mask allocation is more effective in reducing the probability of epidemic as well as epidemic size compared to random allocation. The result echoes the previous findings that mitigation strategies should differ based on the stage of the spreading process, focusing on source control before the epidemic emerges and on self-protection after the emergence. 

\end{abstract}

\maketitle


\section{\label{sec:intro} Introduction}
The COVID-19 pandemic has spread across the globe for over two years, impacting economies and, as of January 2023, had claimed over 6.7 million lives \cite{noauthor_coronavirus_nodate}. 
SARS-CoV-2 is the virus that causes COVID-19.
Human-to-human transmission of SARS-CoV-2 is mainly through coughing, sneezing, and even talking or singing, which spread respiratory droplets or aerosols, with sizes varying from visible to microscopic \cite{howard_evidence_2021, lotfi_covid-19_2020, burke_active_2020, chan_familial_2020, huang_clinical_2020}. 
 When the virus-containing droplets or aerosol particles are exhaled from the infected source, they move forward to a certain distance based on their sizes \cite{noauthor_modes_nodate, shafaghi_effect_2020}. 
A healthy person can be exposed if he inhales droplets or aerosol particles exhaled by an infected person nearby \cite{noauthor_modes_nodate}.
 Scientific research suggests that controlling the COVID-19 pandemic entails both top-down systemic interventions and bottom-up collective changes in public behavior \cite{sohrabi_world_2020, bavel_using_2020, wu_new_2020}.
WHO continues to recommend droplet and contact precautions as non-pharmacological intervention advice for the public, such as mask wearing and reducing social gatherings, etc \cite{noauthor_when_nodate, noauthor_advice_nodate}. 

Compelling data and experimental studies on humans and manikins demonstrate that masking is an effective tool in mitigating SARS-CoV-2 airborne and droplets transmission \cite{sande_professional_2008, leung_respiratory_2020,fischer_effectiveness_2020, davies_testing_2013, lai_effectiveness_2012, dharmadhikari_surgical_2012, patel_respiratory_2016, noauthor_Silk_nodate, duncan_protective_2021, brooks_effectiveness_2021,jain_efficacy_2020, froese_mask_2022,jung_comparison_2014, darby_covid-19_2021,eikenberry_masks, clapham_face_2021, lyu_community_2020}.
The clinical efficacy of a face mask is determined by the filtration efficacy of the material, fit of the mask, and compliance to wearing the mask \cite{sande_professional_2008, darby_covid-19_2021, shafaghi_effect_2020}. 
Two factors are usually considered and tested when assessing the overall clinical efficacy of masks: inward and outward protection efficiency \cite{brooks_effectiveness_2021, jain_efficacy_2020, howard_evidence_2021, jung_comparison_2014, shafaghi_effect_2020}.
The capability of masks that reduces the outward emissions of micron-scale droplets and aerosol particles exhaled by infected persons is known as outward protection, also termed source control \cite{sande_professional_2008, jain_efficacy_2020, jung_comparison_2014, darby_covid-19_2021, brooks_maximizing_2021}.
\cite{noauthor_Silk_nodate} finds that double-masking increases the source control capability of a surgical mask. 
 \cite{darby_covid-19_2021} shows that surgical and cloth masks provide less outward protection potentially due to the weaker seal around these masks compared to N95 respirators. The pressurized droplets or aerosol particles are likely to escape directly from the gap between the mask and the human face due to poor fit. 
Inward protection,  also known as wearer protection, is the capability of masks to act as a barrier to protect the uninfected wearers from respiratory droplets and aerosol particles, to penetrate through and land on exposed mucous membranes of the eye, nose, and mouth \cite{sande_professional_2008, davies_testing_2013, lai_effectiveness_2012, jain_efficacy_2020, tong_3d_2021}.
\cite{yang_mask-wearing_2011, bahl_airborne_2022} finds that cloth masks have limited inward protection in healthcare settings where viral exposure is high compared to surgical masks and N95 respirators.

While it is well-studied that masks {\it qualitatively} mitigate viral spread by limiting the transmission of respiratory droplets,
 many important questions about the {\it quantitative} impact of masks remain open \cite{kabir_prosocial_2021,costantino_impact_2021, eikenberry_masks}. For instance, how many individuals need to wear a mask to prevent future outbreaks \cite{ eikenberry_masks}? When there are not enough high-quality masks (e.g., N95 masks) for all individuals \cite{health_faqs_2021}, how should other types of mass-produced masks (e.g., surgical or cloth masks) be allocated within the population \cite{worby_face_2020}? What types of mask attributes are most desirable in preventing future outbreaks or controlling ongoing pandemics (e.g. source control or self-protection) \cite{lindsley_comparison_2021, sterr_medical_2021,koh_outward_2021}? 

This work aims to answer the above questions from a principled, mathematical lens. We propose a novel agent-based model of viral spread on networks wherein individuals wear different types of masks. In particular, we study models incorporating multiple types of masks; prior work considers scenarios, where individuals either wear a mask or do not wear a mask \cite{eikenberry_masks, worby2020facemask, javid2020impact, kot2020critical, johansson2020masking, tracht_masks, kai2020universal, catching2021examining, silva2020covid, tian2021analysis, Yagan2021Modeling, sridhar2021leveraging, lee2020epidemic, catching_examining_2021}. Our contributions are twofold. First, we derive {\it analytical predictions} for three important epidemiological quantities: the probability of emergence, the epidemic threshold (also known as the basic reproduction number $R_0$), and the expected epidemic size. Specifically, we show how these quantities depend on the structure of the contact network, the properties of the viral spread, and the distribution of masks within the population. Our results are established by leveraging the theory of multi-type branching processes \cite{AthreyaNey, harris_branching_processes}. Moreover, we show through extensive simulations that the analytical predictions we derive are in good agreement with empirical results. 

We then explore a variety of mask-wearing scenarios relevant to the ongoing pandemic, focusing in particular on how the probability of emergence (PE) and expected epidemic size (ES) are affected. First, we quantify the trade-offs between using superior vs. inferior masks (e.g., surgical vs. cloth) when all individuals wear one of either type of mask; naturally, we find that both the PE and ES are reduced when the fraction of superior masks increases in the population. We then consider scenarios in which individuals either wear superior masks, inferior masks, or no masks. We find that increasing the fraction of superior masks significantly decreases the fraction of infected non-mask-wearers when the fraction of non-mask-wearers is small ($<10\%$). Interestingly, when the fraction of non-mask-wearers is larger ($>20\%$), increasing the fraction of superior masks does not significantly mitigate the infections in the non-mask-wearing population. This suggests that mask-wearing strategies are more effective when a larger fraction of the population wears inferior masks, as opposed to a smaller fraction of the population wearing superior masks. 
Next, we study trade-offs between masks that are ``inward-good" (i.e., good at blocking respiratory droplets from the outside but poor at limiting transmission from the mask-wearer to the outside), and those that are ``outward-good" (i.e., good at limiting transmission from the mask-wearer to the outside but poor at blocking respiratory droplets from the outside) \cite{koh_outward_2021, pan2021masks, jain_efficacy_2020, sande_professional_2008}. 
We find that both mask types are useful but at different stages of viral propagation. In particular, outward-good masks are most helpful in preventing the emergence of an epidemic, while inward-good masks are best for reducing the infections of an already ongoing epidemic. 
Lastly, we look into mask assignment depending on node degree, by assigning outward-good masks to top $x\%$ high (low) degree nodes and inward-good masks to the rest of the nodes. We find a scenario in which high-degree nodes wear inward-good masks and low-degree nodes wear outward-good masks is efficient in reducing the probability of emergence and the opposite allocation scheme is more helpful in controlling the epidemic size extension after the epidemic forms. This result reconfirms that we need to treat the two stages of virus spread (i.e. before and after the epidemic exists) with different mitigation strategies. It also indicates that high-degree nodes and low-degree nodes play different roles in the epidemic process.  The most powerful factor leading to a pandemic as well as extending the pandemic is high degree nodes. However, before the epidemic starts, removing the additional infecting paths from low-degree initiator to susceptible high-degree nodes is critical in preventing the epidemic from happening. After the epidemic forms, protecting susceptible low-degree nodes from infected high-degree nodes is more important in suppressing the propagation of the epidemic. 

While our results are motivated by mask-wearing in a pandemic, our model can be applied more broadly to other mitigation strategies. For example, our results on inward and outward efficiencies suggest that when we think of a mitigation strategy for a pandemic, we should  consider the current stage of the spread: early on, it is most important to limit people spreading it to others (this can be achieved through social distancing, for instance), while later it becomes more important to protect individuals from getting the virus. This insight can help with strategies for prioritizing vaccines, limiting gatherings, etc. On a more technical note, our model generally captures heterogeneities in the capability of a node to be affected by or to spread a virus. While one interpretation of node-level heterogeneity is the type of mask used, it can also capture the effects of vaccinations or community-based interactions (i.e., if individuals tend to interact within their own community rather than within others, this may result in higher transmissibility between two individuals of the same community).

The structure of this paper is as follows. In Section \ref{sec:epidemic_models}, we provide an overview of related epidemic models and introduce a formal description of a model for viral spread in the presence of masks of various types. Section \ref{sec:analysis} contains our theoretical analysis, where we derive expressions for the probability of emergence, the epidemic threshold, and the expected size of the epidemic. Our theoretical results are verified in Section \ref{sec:numerical_results}, where we explore the implications of the multi-type mask model through simulations. Finally, we conclude and discuss future avenues of research in Section \ref{sec:conclusion}. 
Appendix Section \ref{apdix:Preliminaries} to \ref{apdix:convergence_phasetransition_ES} present preliminaries and more technical details for analytical results.
Appendix Section \ref{apdix:additional_validation} provides additional experiments validating analytical results.
Appendix Section \ref{apdix:sa_network_structures} presents a sensitivity analysis to explore the impact of the network structure with four different experiment settings, including replacing the current network model with random networks with clustering \cite{newman_random_2009}, as well as a real-world dataset. 

\begin{figure*}[ht]
    \centering
    \subfigure[]{
    \label{fig:starting}
    \includegraphics[width=0.22\textwidth, angle=90]{ 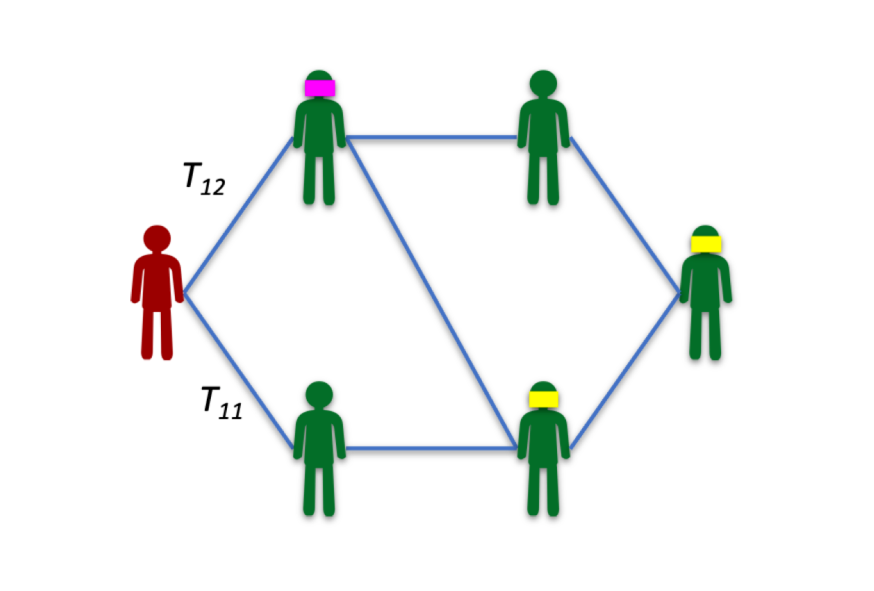}}
    \subfigure[]{
    \label{fig:first_layer}
    \includegraphics[width=0.22\textwidth, angle=90]{ 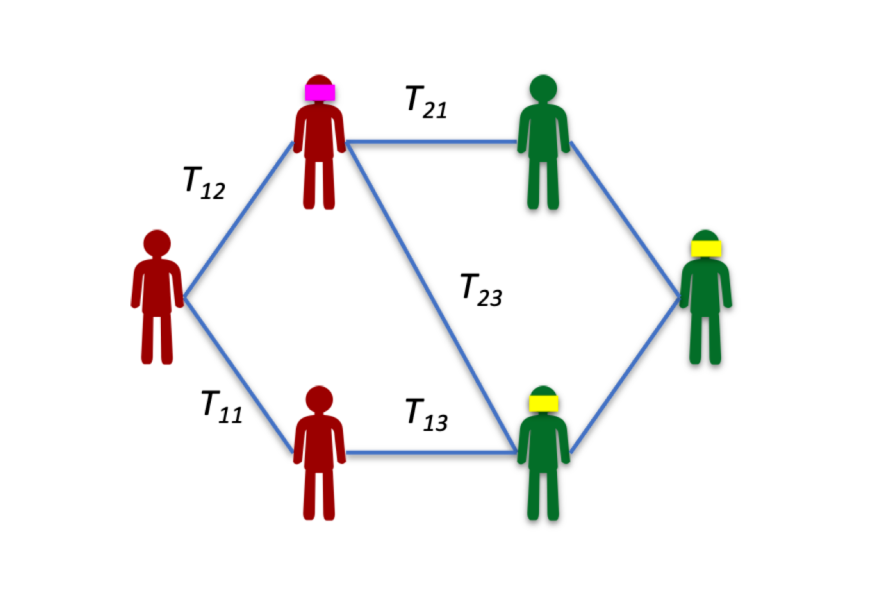}}
    \subfigure[]{
    \label{fig:end_1}
    \includegraphics[width=0.22\textwidth, angle=90]{ 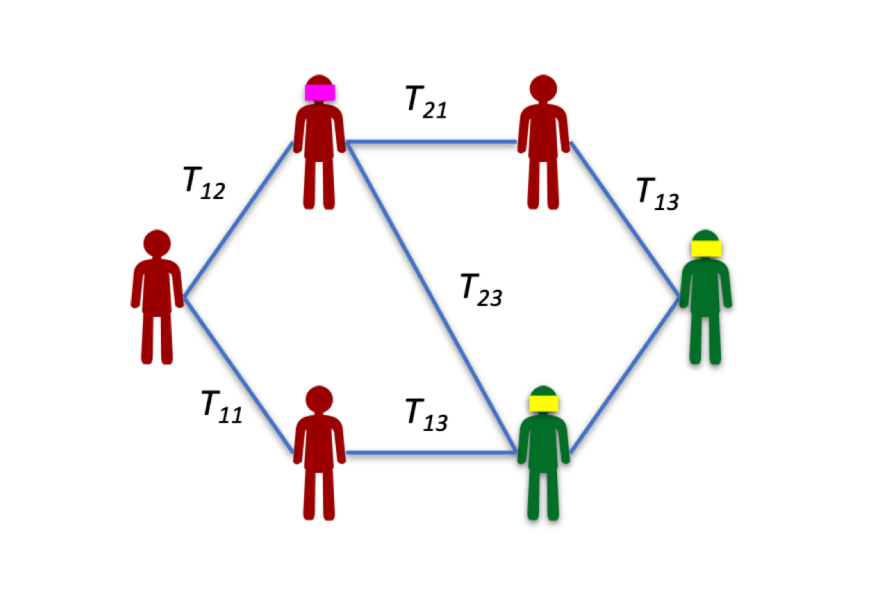}}
    \subfigure[]{
    \label{fig:end}
    \includegraphics[width=0.22\textwidth, angle=90]{ 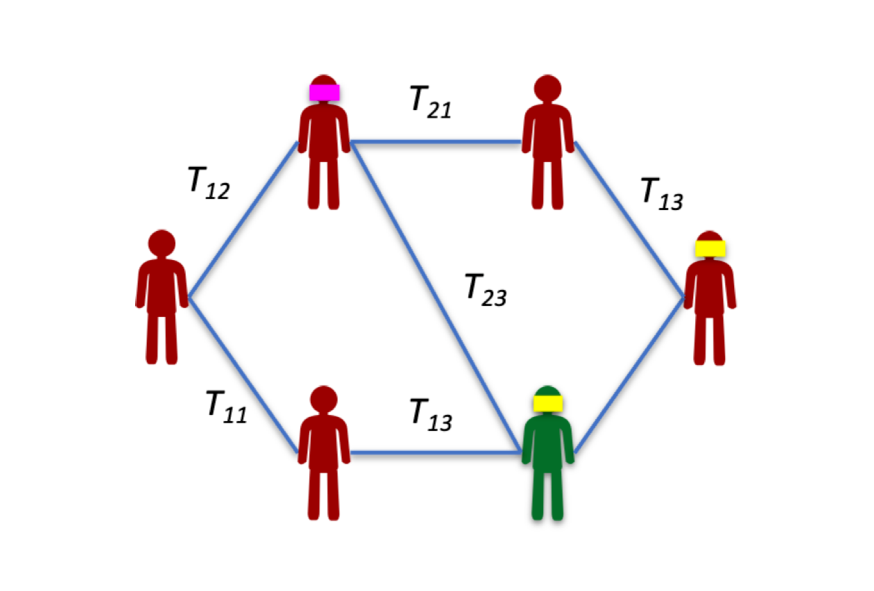}}
    
    \caption{ A simple demonstration of the spreading process with a population of 6. Nodes wearing pink masks is type-1, and yellow masks is type-2. Red color means infected, while green means healthy. 
    (a) Before the transmission starts, the status of mask-wearing for the population is decided. The transmissibilities of the seed to two of his contacts are set according to their mask-wearing status. There is one initially infected seed (layer 0). 
    (b) The seed successfully infected his neighbors (layer 1) with different transmissibilities. 
    (c) Layer 1 nodes only infected the node that doesn't wear masks (upper right node). 
    (d) The upper right node continues the infection to the last layer but didn't succeed.}
    \label{fig:schematic}
    \vspace{-3mm}
\end{figure*}
\section{\label{sec:epidemic_models} Epidemic Models}
The literature on epidemic modeling generally falls into one of two categories: ordinary differential equation (ODE) models, and agent-based stochastic models. In typical ODE models, the evolution of the fraction of various types (e.g., susceptible individuals, infected individuals) is studied. The governing equations for the epidemic dynamics are derived from laws of mass action that are based on the way individuals interact and spread the virus (see, e.g., \cite{brauer2012mathematical}). A number of recent works have used the ODE approach to model viral spread with masks. Tracht et al.  \cite{tracht_masks} first proposed an ODE model incorporating mask-wearing. Motivated by challenges caused by the COVID-19 pandemic, many authors have subsequently elaborated on the model of \cite{tracht_masks}, addressing for instance the effect of asymptomatic infections and mask allocation \cite{eikenberry_masks, worby_face_2020, javid2020impact, kot2020critical, johansson2020masking}. However, a common criticism of the ODE approach is that in modeling {\it population-level} phenomena, it fails to consider the possibly complex ways in which {\it individuals} interact with each other.

Agent-based stochastic models address this gap by studying how individual-level interactions facilitate viral spread. Such models reveal the rich interplay between interaction patterns and viral spread. Various agent-based models incorporating mask-wearing have been recently proposed \cite{kai2020universal, catching2021examining, silva2020covid, tian2021analysis, Yagan2021Modeling, sridhar2021leveraging, lee2020epidemic}. The drawback of stochastic models is that they are often challenging to simulate for realistic parameters. To overcome this issue, a significant line of work on agent-based models focuses on deriving {\it analytical predictions} for key epidemiological quantities that accurately describe viral spread in large contact networks; the works \cite{tian2021analysis, Yagan2021Modeling, sridhar2021leveraging, lee2020epidemic} adopt this approach to quantify the effect of mask-wearing. This is also the focus of the present paper. We emphasize that the literature discussed above studies the effect of a {\it single type} of mask in the population. A key novelty of our work is that we study the effect of {\it multiple types} of masks which helps reveal the trade-offs between the inward and outward efficiencies of the masks involved and, more broadly, sheds light on the effectiveness of various mitigation strategies.  

In the remainder of this section, we describe stochastic models of viral spread in more detail. In Section \ref{subsec:simple}, we review the basic stochastic model. We then discuss a new stochastic model with multi-type masks in Section \ref{subsec:multi-type}. In section \ref{subsec:network-model}, we provide an overview of the network model we adopt for our underlying contact network.

\subsection{An agent-based stochastic model on networks}
\label{subsec:simple}

In his seminal work \cite{Newman_2002}, Newman considered an SIR model of viral spread over a contact network. Initially, there is a single infected individual in the population. When a susceptible individual and an infected individual interact, the infected individual transmits the virus to the susceptible individual with probability $T$, where $T$ is called the {\it transmissibility} of the virus. After a fixed or random amount of time, infected individuals recover and can no longer transmit the virus. Through branching process techniques, Newman derived expressions for two key epidemiological quantities: the probability that an epidemic emerges (PE) and the expected size of the epidemic (ES). Importantly, his results revealed the crucial role that the structure of the contact network plays in the behavior of the PE and ES. Although the viral dynamics may appear simple, Newman's model can capture complex viral transmission and recovery mechanisms through the appropriate choice of $T$ \cite[Equations (2)-(6)]{Newman_2002}. Several authors have generalized Newman's model to account for various agent-level heterogeneities \cite{sridhar2021leveraging, Yagan2021Modeling, tian2021analysis, lee2020epidemic, yagan2013conjoining,eletreby2020effects, alexander2010risk, allard2009heterogenous}.

\subsection{Viral spread with multi-type masks}
\label{subsec:multi-type}

In this work, we consider a generalization of Newman's basic framework called the {\it multi-type mask model}. Motivated by mask-wearing behaviors in response to the COVID-19 pandemic, we assume that there are $M$ types of individuals, each wearing a different type of mask. For notational convenience, we shall say that an individual is of type-$i$ if they wear a type-$i$ mask (here, $1 \le i \le M$). We assume that the probability of transmission between individuals varies depending on the type of mask each individual wears. Specifically, the probability that a virus is eventually transmitted from a type-$i$ infective to a type-$j$ susceptible is $\mathbf{T}_{ij}$, where $\mathbf{T}$ is an $M \times M$ transmissibility matrix. 

Typically, masks are characterized in terms of their inward and outward efficiencies (see, e.g., \cite{pan2021masks, brooks_effectiveness_2021,jain_efficacy_2020, howard_evidence_2021, jung_comparison_2014, shafaghi_effect_2020}). 
The inward efficiency is the probability that respiratory droplets will be \textit{blocked} from the outside layer of the mask to the inside; thus, inward efficiency quantifies the protection of the mask against receiving the virus.  The outward efficiency is the probability that respiratory droplets will \textit{kept} from the inside layer of the mask to the outside, quantifying the protection against transmitting the virus. The transmission probability from a type-$i$ individual to a type-$j$ individual is then given by
\begin{equation}
\label{eq:T_inward_outward}
\mathbf{T}_{ij} : = (1 - \epsilon_{\textrm{out},i}) (1 - \epsilon_{\textrm{in},j}) T,
\end{equation}
where $\epsilon_{\textrm{out},i}$ is the outward efficiency of a type-$i$ mask, $\epsilon_{\textrm{in},j}$ is the inward efficiency of a type-$j$ mask, and $T$ is the baseline transmissibility of the virus, i.e., the probability of transmission in the presence of no masks.
We have $\epsilon_{\textrm{out},i}, \epsilon_{\textrm{in},j} \in [0, 1], \forall i, j \in [1, M]$. 
The higher the inward or outward efficiency is, the smaller $\mathbf{T}_{i,j}$ is, and the better quality the mask is.
In Equation (\ref{eq:T_inward_outward}), $(1- \epsilon_{\textrm{out}})$ ($(1 - \epsilon_{\textrm{in}})$, correspondingly) represents the probability that the droplets will \textit{pass} the mask from the inside (outside) layer of the mask to the outside (inside).
 Also Note that $\mathbf{T}$ is not symmetric if the vectors
$(1 - \epsilon_{\textrm{out}})$ and
$(1 - \epsilon_{\textrm{in}})$ are not collinear.
We would like to remark that the multi-type mask model can generally capture the effects of mitigation strategies that introduce node heterogeneity of taking in and spreading out the spreading item. 
For example, in a case where a fraction of the population is vaccinated against the virus, we can consider a new type of \textit{mask} (or, equivalently, a new type of nodes) with appropriate inward and outward efficiency parameters associated with that type.
If in a given context vaccinated individuals gain full immunity against the virus, we can capture that by setting $\epsilon_{out,v}=\epsilon_{in, v} = 1$ for nodes of type-$v$ that represents vaccinated individuals.
This setting will ensure that a vaccinated individual can never be infected.


In line with prior literature on stochastic epidemic models, we generate the contact network $\mathcal{G}$ by the {\em configuration model} \cite{molloy1995critical}. 
Namely, we specify a distribution $\{p_k \}_{k \ge 0}$ with support on the non-negative integers, where $p_k$ is the probability that an arbitrary vertex has {\em degree} $k$, i.e., it is connected to $k$ other nodes via an {\em undirected} edge. According to the configuration model, the degrees of vertices in $\mathcal{G}$ are drawn independently from $\{p_k \}_{k \ge 0}$. Equivalently, 
$\mathcal{G}$ is selected uniformly at random from among all graphs satisfying the {degree} distribution $p_k$.
Next, we assume that the $M$ types of masks are randomly distributed amongst vertices in the $\mathcal{G}$. Let $\{m_1, \ldots, m_M\}$ be a distribution over the set $\{1, \ldots, M \}$ where $m_i$ represents the fraction of individuals who wear a mask of type-$i$. We further assume that the type of mask is chosen independently\footnote{The independence assumption is for mathematical tractability, as it implies that the distribution of mask types within a vertex's neighborhood is given by $\{m_i\}_{i = 1}^M$. One can imagine realistic scenarios where this is not the case; for instance, individuals who wear a mask may be more likely to interact with other mask-wearers, rather than non-mask-wearers.} from $\{m_i \}_{i =1}^M$ over all vertices in $G$. See Table \ref{table:parameters} for a succinct description of our model parameters, and Figure \ref{fig:schematic} for an illustration of the spreading process.  For each $i=1, \ldots, M$, our goal is to compute the following quantities of interest: the PE assuming that the initially infected node is of type-$i$, and the ES of the infected type-$i$ individuals. 

\begin{table}
\centering
\begin{tabularx}{0.9 \columnwidth} {
| >{\centering \arraybackslash}X
| >{\raggedright \arraybackslash}X|}
\hline
{\bf Quantity} & {\bf Description} \\
\hline
$M$ & Number of mask types \\
\hline 
$m_i, 1 \le i \le M$ & The probability that a given individual is type-$i$ \\
\hline
$T$ & Baseline virus transmissibility \\
\hline
$\mathbf{T}_{ij}$, $1 \le i,j \le M$ & The probability that an infected type-$i$ individual infects a type-$j$ neighbor ($\mathbf{T}$ is the transmissibility matrix) \\
\hline
$\epsilon_{\textrm{out},i}$, $\epsilon_{\textrm{in}, i}$ & Outward and inward efficiencies of type-$i$ mask\\
\hline
$\mathcal{G}$ & Graph representing the contact network \\
\hline
$\{p_k \}_{k \ge 0}$ & Degree distribution for $G$ generated via the configuration model. \\
\hline 
\end{tabularx}
\caption{Description of parameters in the multi-type mask model.}
\label{table:parameters}
\end{table}

The above modification of Newman's model for the mask-wearing setting has garnered recent interest in the literature. In previous work, we studied the special case where $M = 2$: individuals either wear a mask or do not \cite{tian2021analysis}. We also derived expressions for PE and ES in this setting. Lee and Zhu \cite{lee2020epidemic} simultaneously studied the same model, and obtained results for the ES in the setting $M = 2$. The results of both \cite{tian2021analysis} and \cite{lee2020epidemic} follow as a special case of our more general model. We also mention the work of Allard et al. \cite{allard2009heterogenous}, which is especially relevant. They study a bond percolation problem over multi-type networks, which can be viewed as a more generic version of the mask model we study here. A key difference between our work and theirs is that the distribution of mask types and the network formation are {\it independent} in our model, whereas Allard et al. consider a framework where the probability of connectivity among nodes also depends on the node types, resulting a joint generation of node types and network structure. This renders their results  harder to interpret, albeit more general. Our formulation yields simpler formulae that clearly illustrate how the structural aspects of the network, mask properties, and viral transmission dynamics interact to derive the PE and ES. In addition, our work also validates the theoretical analysis through extensive simulations and provides insights into the trade-offs between inward-good and outward-good masks.  
Finally, our framework still has the flexibility to correlate the mask types with the network structure by  modifying the mask allocation strategy based on node degrees. In the Results section, we demonstrate this and show that  allocating different types of masks based on node degree can help further reduce the probability and  size of epidemics.

\subsection{Network model}
\label{subsec:network-model}
To model the underlying contact network, we utilize random graphs with random degree distribution generated by the configuration model \cite{newman2001random, molloy1995critical}. 
The configuration model generates random graphs with specified degree sequences sampled from a degree distribution. Let $\mathcal{G}$ denote the underlying contact network defined on the node set  $ \mathbf{N} =  \{1, ..., n\}$. The structure of $\mathcal{G}$ is defined through its degree distribution $\{p_k, k = 0, 1, ...\}$, where $p_k$ is the probability that an arbitrary node in network $\mathcal{G}$ has degree $k$. 
We generate a degree sequence from a \textit{well-behaved} degree distribution, i.e., distribution with  moments of arbitrary order being \textit{finite} \cite{yagan2013conjoining, Newman_2002, newman_configuration_2018}. 
Our analytical solutions are valid for such well-behaved distributions, e.g., Poisson 
degree distributions, power-law degree distributions with exponential cut-off, etc.
On the other hand, it is worth noting that if the second moment of the degree distribution is finite when $n$ approaches infinite, the expected clustering coefficient of a typical vertex approaches zero. This indicates that the graph is locally tree-like. 
In the context of the configuration model, the degree distribution of a randomly chosen neighbor of a randomly chosen vertex is denoted by $ \left\{\hat{p}_{k}\right. k = 1, 2, ... \}$, and is given by $\hat{p}_{k}=k p_{k} /\langle k\rangle$ for $k = 1, 2, ...$, where $\langle k\rangle$ denotes the mean degree (i.e., $\langle k\rangle = \sum_k k p_k$).

\section{\label{sec:analysis} Analytical Results}
In this section, we present the derivation of the probability of emergence (PE), the epidemic threshold ($R_0$), and the expected epidemic size (ES) in the multi-type mask model for an arbitrary integer $M>1$. Formally, {\it emergence} is defined to be the event where the virus infects an infinite or unbounded number of vertices in the network. It is the complement of the {\it extinction} event, in which the virus dies out after infecting a finite number of individuals. In Section \ref{subsec:pe}, we compute the through an approach based on probability generating functions (PGFs). 
The works \cite{Newman_2002, eletreby2020effects, tian2021analysis, alexander2010risk} use this method to derive the PE for related epidemic models, among which \cite{tian2021analysis} provide a similar analysis for the $M=2$ case partially relying on the conclusion from \cite{eletreby2020effects}. In our work, we present a direct derivation for any $M>1$ case.
Preliminaries for PGFs and configuration models can be found in Appendix Section \ref{apdix:Preliminaries}.

In Section \ref{subsec:R0}, we study the epidemic threshold $R_0$, also known as the basic reproduction number. When $R_0 \leq 1$, the epidemic dies out almost surely and when $R_0 > 1$, there is a positive probability that the epidemic emerges. Generally, $R_0$ is  interpreted as the mean number of secondary infections caused by a given infective (see, e.g., \cite{Newman_2002}), but our results show that the true picture is more subtle than that. Indeed, using classical results from multi-type branching process theory \cite{AthreyaNey, harris_branching_processes}, we show that $R_0$ is the spectral radius of a matrix that depends on $\{m\}_{i = 1}^M, \{T_{ij}\}_{1 \le i,j \le M}$ as well as the first and second moments of the degree distribution. 

Finally, in Section \ref{subsec:es}, we study the ES. Specifically, we show how to compute the expected fraction of type-$i$ individuals, {\it conditioned} on the emergence of the epidemic. To do so, we leverage the method of Gleeson and coauthors \cite{gleeson2007seed, gleeson2008cascades}, which were  recently used to compute the ES in models of viral spread \cite{eletreby2020effects, tian2021analysis}. 



\subsection{Probability of emergence}
\label{subsec:pe}

Suppose that a type-$i$ infective -- named $v$ for convenience -- has $k_j$ susceptible neighbors of type $j$, for $1 \le j \le M$. Let $X_j$ be the number of neighbors of type $j$ who are eventually infected by $v$, so that $X_j \sim \mathrm{Binomial}(k_j, \mathbf{T}_{ij})$. Moreover, $X_1, \ldots, X_M$ are independent. Conditioned on the neighborhood profile $k_1, \ldots, k_M$, the PGF of the number of infections of each type  caused directly by $v$ is 
$$
\mathbb{E} \left [ s_1^{X_1} \ldots s_M^{X_M}\mid k_1, \ldots, k_M \right ] = \prod\limits_{j = 1}^M \left( 1 - \mathbf{T}_{ij} + \mathbf{T}_{ij} s_j \right)^{k_j}.
$$
Our next step is to condition on the {\it total} number of neighbors of $v$ rather than the number of neighbors of each type. Note that if we are given $k : = k_1 + \ldots + k_M$, the tuple $(k_1, \ldots, k_M)$ is drawn from the Multinomial distribution. Namely, the probability of observing a given instantiation $(k_1, \ldots, k_M)$ is equal to 
$$
{k \choose k_1, \ldots, k_M} m_1^{k_1} \ldots m_M^{k_M}.
$$
The PGF of the number of neighbors of all types directly caused by $v$ conditioned on $k$ can therefore be written as
\begin{align}
& \mathbb{E} \left[ s_1^{X_1} \ldots s_M^{X_M} \mid k \right] \nonumber  \\ 
& \hspace{0.5cm} = \hspace{-0.4cm} \sum\limits_{k_1 +\ldots + k_M = k} {k \choose k_1, \ldots, k_M} \prod\limits_{j = 1}^M \left( m_j ( 1 - \mathbf{T}_{ij} + \mathbf{T}_{ij} s_j ) \right)^{k_j} \nonumber \\
\label{eq:pe_basic_pgf}
& \hspace{0.5cm} = \left( \sum\limits_{j = 1}^M m_j ( 1 - \mathbf{T}_{ij} + \mathbf{T}_{ij} s_j ) \right)^k.
\end{align}
Finally, to remove the conditioning on the number of neighbors, $k$, we take an expectation over the degree distribution of $v$. If $v$ is type-$i$, the PGF of the number of secondary infections of each type is given by 
\begin{align*}
\gamma_i(s_1, \ldots, s_M) & : = \sum\limits_{k = 0}^\infty p_k \left( \sum\limits_{j = 1}^M m_j (1 - \mathbf{T}_{ij} + \mathbf{T}_{ij} s_j ) \right)^k \\
& = g \left( \sum\limits_{j = 1}^M m_j (1 - \mathbf{T}_{ij} + \mathbf{T}_{ij} s_j ) \right).
\end{align*}
where $g(x)$ is the PGF of the degree distribution of an arbitrary vertex in the configuration model given by
$$
g(x) : = \sum\limits_{k = 0}^\infty p_k x^k.
$$
For notational convenience, we combine the PGFs of all types into a single vectorized PGF $\boldsymbol{\gamma} : = \{\gamma_i \}_{i = 1}^M$.

On the other hand, if $v$ is a {\it later-generation} infective, the number of its children follows the excess degree distribution. Hence the PGF of the number of secondary infections of each type is given by 
$$
\Gamma_i(s_1, \ldots, s_M) := G \left( \sum\limits_{j = 1}^M m_j (1 - \mathbf{T}_{ij} + \mathbf{T}_{ij} s_j ) \right).
$$
where $G(x)$ is the PGF of the excess degree distribution with the form of:
$$
G(x) : = \sum\limits_{k = 0}^\infty \frac{k p_k }{ \langle k \rangle} x^{k-1}.
$$
We also define the vectorized PGF $\boldsymbol{\Gamma} : = \{ \Gamma_i \}_{i = 1}^M$. 

With these PGFs in hand, we can compute the probability of extinction. Formally, for $1 \le i \le M$ and a positive integer $n$, let $P_i^{(n)}$ denote the probability that the epidemic dies out by generation $n$ (that is, distance $n$ from the infection source), where the initial infective is of type $i$. Classical results from branching process theory \cite[Chapter V.1]{AthreyaNey} imply that
$$
(P_1^{(n)}, \ldots, P_M^{(n)} ) = (\boldsymbol{\gamma} \circ \underbrace{\boldsymbol{\Gamma} \circ \ldots \circ \boldsymbol{\Gamma}}_\text{$n-1$ times}) (0, \ldots, 0).
$$
The probabilities of  extinction can be found by taking the limit $n \to \infty$. Formally, define $\boldsymbol{\Gamma}^{(n)}$ to be the $n$-fold composition of $\boldsymbol{\Gamma}$. Assuming there is a well-defined limit as $n \to \infty$, $\boldsymbol{\Gamma}^{(n)}(0,\ldots, 0) \to (Q_1, \ldots, Q_M)$ which is the unique vector satisfying
$$
\boldsymbol{\Gamma}(Q_1, \ldots, Q_M) = (Q_1, \ldots, Q_M). 
$$
The probabilities of eventual extinction are therefore given by 
\begin{equation}
\label{eq:final_PE}
(P_1, \ldots, P_M) = \boldsymbol{\gamma}(Q_1, \ldots, Q_M).
\end{equation}
Since the above describes the extinction event when the initial infective is type-$i$, the PE is $1 - P_i$. 

To formally justify taking the limit $n \to \infty$, we need to check that the multi-type branching process with PGF $\mathbf{\Gamma}$ is {\it positive regular} and {\it non-singular} \cite{AthreyaNey, harris_branching_processes}. The process is singular if and only if each type has exactly one secondary infection. Our model is clearly non-singular since each neighbor of an infective is independently infected. A sufficient condition for our process to be positive regular is if there is a positive probability that a type-$i$ individual can infect a type-$j$ individual, for any $1 \le i, j \le M$. This is indeed the case if we assume that $m_i > 0$ and $\mathbf{T}_{ij} > 0$ for all $1 \le i,j \le M$. Both assumptions are expected to hold in practice as the former condition states that there is a positive fraction of the population wearing each type of mask, and the latter states that there is a positive probability of transmission between any two neighboring individuals. 

\subsection{Epidemic threshold}
\label{subsec:R0}

In the special case where $M = 1$ (e.g., when no one wears a mask), it is well known that there exists a phase transition in the PE based on the {\em basic reproduction number}, $R_0$, defined as the mean number of {\em secondary} infections in a {\em naive} population. Put differently, $R_0$ is the expected number of new infections generated by a newly infected node in a population where all individuals are susceptible.
It is known that if $R_0$ is greater than one then the PE is positive, i.e., epidemics can take place. When $R_0 \le 1$ on the other hand, the PE is zero \cite{Newman_2002} meaning that there is zero chance for a spreading process to reach a positive fraction of the population. Beyond marking a phase transition, the metric $R_0$ measures, in a sense, the speed at which the epidemic grows and is often used by policymakers when deciding on mitigation strategies. Thus, it is of significant importance to characterize $R_0$ for the multi-type mask model.

Firstly, notice from the computations in Section \ref{subsec:pe} (in particular, from \eqref{eq:final_PE}) that the PE is zero ($P_1 = \ldots = P_M = 1$) if and only if $Q_1 = \ldots = Q_M = 1$. Hence we focus on the phase transition for extinction/emergence of the multi-type branching process with PGF $\mathbf{\Gamma}$. Let $A \in \mathbb{R}^{M \times M}$ be a non-negative matrix such that $A_{ij}$ is the expected number of type-$j$ children\footnote{For an infected node, all the new infected nodes directly generated by this node are called its children, also known as its later-generation.} of a type-$i$ later-generation infective. Define $R_0 : = \rho(A)$ to be the spectral radius of $A$, and recall that the multi-type branching process with PGF $\boldsymbol{\Gamma}$ is positive regular and non-singular (see the discussion at the end of Section \eqref{subsec:pe}). It is a well-known property of multi-type branching processes that if $R_0 > 1$ we have $P_i < 1$ for all $1 \le i \le M$, while if $R_0 \le 1$ we have $P_i = 0$ for all $1 \le i \le M$; e.g., see \cite{harris_branching_processes, AthreyaNey}. 

We proceed by computing the entries of $A$ to illustrate the dependence of $R_0$ on the graph topology, the mask parameters and the viral transmissibilities. To this end, note that $A_{ij}$ is equal to the expected number of type-$j$ children of a later-generation infective times $T_{ij}$. The expected number of children (computed over all types) is given by 
$$
G'(1) = \sum\limits_{k = 1}^\infty \frac{ k(k-1) p_k }{\langle k \rangle} = \frac{ \langle k^2 \rangle - \langle k \rangle }{\langle k \rangle},
$$
where $\langle k \rangle$ and $\langle k^2 \rangle$ are the first and second moments of the degree distribution, respectively. Since the probability that a given child is type-$j$ is $m_j$, we have
$$
A_{ij} = \left( \frac{ \langle k^2 \rangle - \langle k \rangle }{\langle k \rangle} \right) \mathbf{T}_{ij} m_j.
$$
This leads to the matrix representation
$$
A = \frac{ \langle k^2 \rangle - \langle k \rangle }{\langle k \rangle} \mathbf{\mathbf{T}m},
$$
where the $(i,j)^{\textrm{th}}$ entry of the matrix $\mathbf{T}$ is $\mathbf{T}_{ij}$ and $\mathbf{m}$ is a diagonal matrix with $(i,i)^{\textrm{th}}$ entry being $m_i$. Putting everything together, we have
\begin{equation}
\label{eq:R0}
R_0 = \frac{ \langle k^2 \rangle - \langle k \rangle }{\langle k \rangle} \rho( \mathbf{\mathbf{T}m})
\end{equation}

We next study a simpler version of \eqref{eq:R0} in a special case of interest. Typically, masks are characterized in terms of their inward and outward efficiencies (see, e.g., \cite{pan2021masks}). The inward efficiency is the probability that respiratory droplets will pass from the outside layer of the mask to the inside; thus, inward efficiency quantifies the protection of the mask against receiving the virus.  The outward efficiency is the probability that respiratory droplets will pass from the inside layer of the mask to the outside, quantifying the protection against transmitting the virus. The transmission probability from a type-$i$ individual to a type-$j$ individual is then given by
$$
\mathbf{T}_{ij} : = (1 - \epsilon_{\textrm{out},i}) (1 - \epsilon_{\textrm{in},j}) T,
$$
where $\epsilon_{\textrm{out},i}$ is the outward efficiency of a type-$i$ mask, $\epsilon_{\textrm{in},j}$ is the inward efficiency of a type-$j$ mask, and $T$ is the baseline transmissibility of the virus, i.e., the probability of transmission in the presence of no masks. For notational convenience, let $\boldsymbol{\epsilon}_{out}$ be the $M$-dimensional vector where the $i^{\textrm{th}}$ entry is $\epsilon_{\textrm{out},i}$; similarly define the vector $\boldsymbol{\epsilon}_{in}$. Then, using \eqref{eq:R0}, we have
$$
R_0 = \left( \frac{ \langle k^2 \rangle - \langle k \rangle }{\langle k \rangle} \right) T \rho(( \boldsymbol{1 - \epsilon}_{out}) ( \boldsymbol{1 - \epsilon}_{in})^\top \mathbf{m} ).
$$
Since $(\boldsymbol{1 - \epsilon}_{out}) ( \boldsymbol{1 - \epsilon}_{in})^\top \mathbf{m}$ is a rank one matrix, it has only one non-zero eigenvalue which is given by 
$$
 (\boldsymbol{1 - \epsilon}_{in})  
 ^\top
 \mathbf{m}
 ( \boldsymbol{1 - \epsilon}_{out}) = \sum\limits_{i = 1}^M m_i (1 - \epsilon_{\textrm{in},i}) (1 - \epsilon_{\textrm{out},i}).
$$
This allows us to conclude that 
\begin{align}
R_0 & = \left( \frac{ \langle k^2 \rangle - \langle k \rangle }{\langle k \rangle} \right) T \sum\limits_{i = 1}^M m_i (1 - \epsilon_{\textrm{in}, i}) (1 - \epsilon_{\textrm{out},i}) \nonumber \\
\label{eq:R0_epsilon}
\end{align}

It is worth noticing that the simplifications that yield to Equation (\ref{eq:R0_epsilon}) do \textit{not}
 apply to all heterogeneous bond percolation models with arbitrary transmissibility matrix $\mathbf{T}$.
 In other words, the basic reproduction number $R_0$ might depend on cross terms $\mathbf{T}_{ij}$ for an arbitrary transmissibility matrix $\mathbf{T}$.
In the specific case studied in this work, the transmissibility matrix $\mathbf{T}$ is given by the inward and outward efficiency parameters of the $M$ mask types, and the basic transmissibility parameter $T$.
This leads all rows of $\mathbf{T}$ being linearly dependent on each other, e.g., the $i^{\textrm{th}}$ row can be found by multiplying the $j^{\textrm{th}}$ row by ${(1 - \epsilon_{\textrm{out}, i})}/{(1 - \epsilon_{\textrm{out}, j})}$.
Put differently, the transmissibility matrix $\mathbf{T}$ resulting from different mask types is rank-1.
It is this special property that allows the simplifications to Equation (\ref{eq:R0_epsilon}).
On the other hand, Equation (\ref{eq:R0_epsilon}) disentangles three different factors that contribute to the spreading processes:
network structure $\left( \frac{ \langle k^2 \rangle - \langle k \rangle }{\langle k \rangle} \right)$, 
viral transmissibility $T$ and
average mask filtration power $\sum\limits_{i = 1}^M m_i (1 - \epsilon_{in, i}) (1 - \epsilon_{out,i})$.
We point out that 
$(1 - \epsilon_{in, i}) (1 - \epsilon_{out,i})$ is a property for each type of masks, 
which also echoes how we evaluate the average mask filtration power in Section \ref{subsec: inout_tradeoff}.

\begin{figure}[h!]
\centering
    \subfigure[$T$ - $m_{\text{no-mask}}$ epidemic boundary]{
    \label{fig:figure2a}
    \includegraphics[width=0.3\textwidth]{ 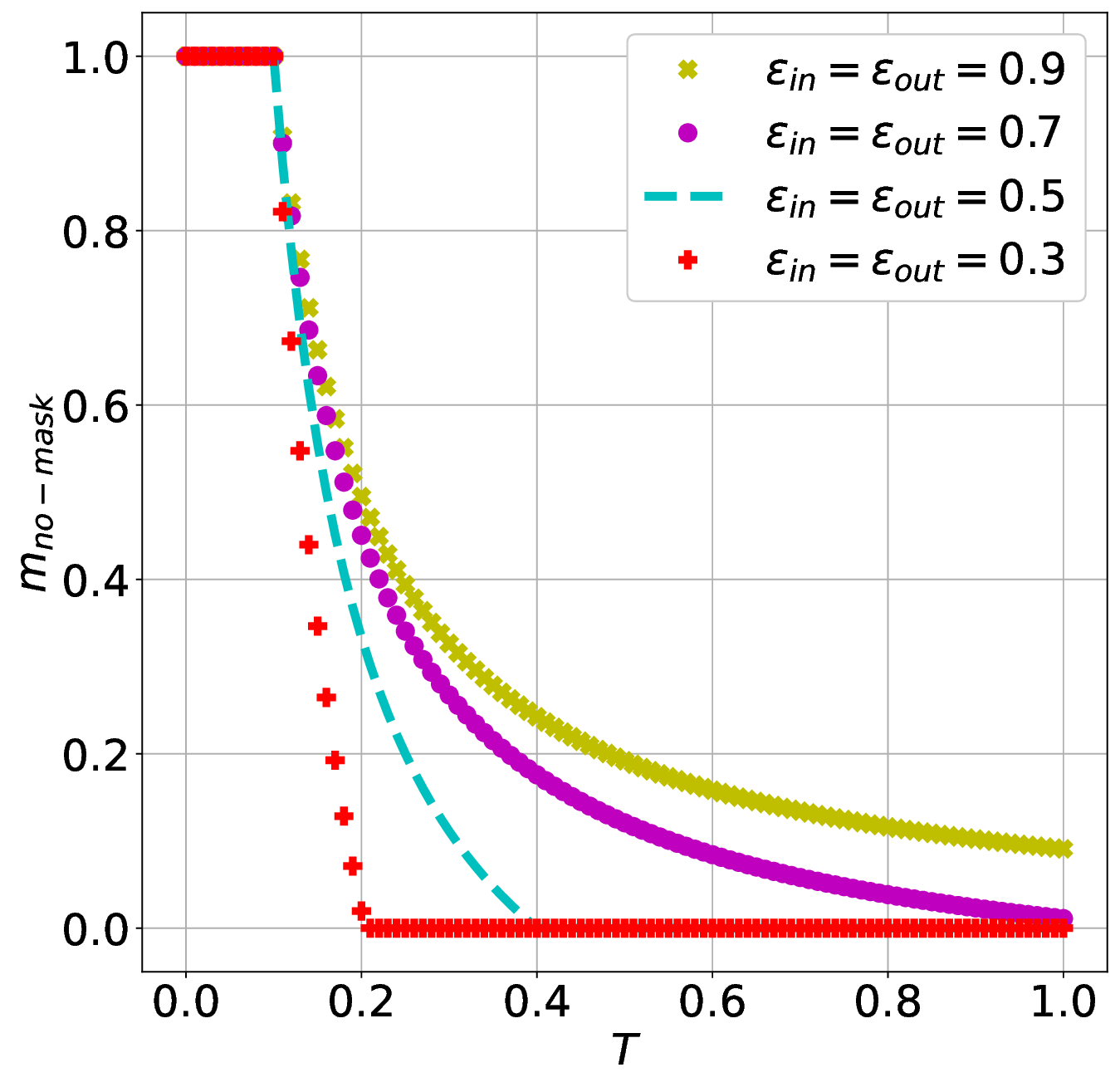}}
    \subfigure[$m_{\text{mask}}$ - $\epsilon$ epidemic boundary]{
    \label{fig:figure2b}
    \includegraphics[width=0.3\textwidth]{ 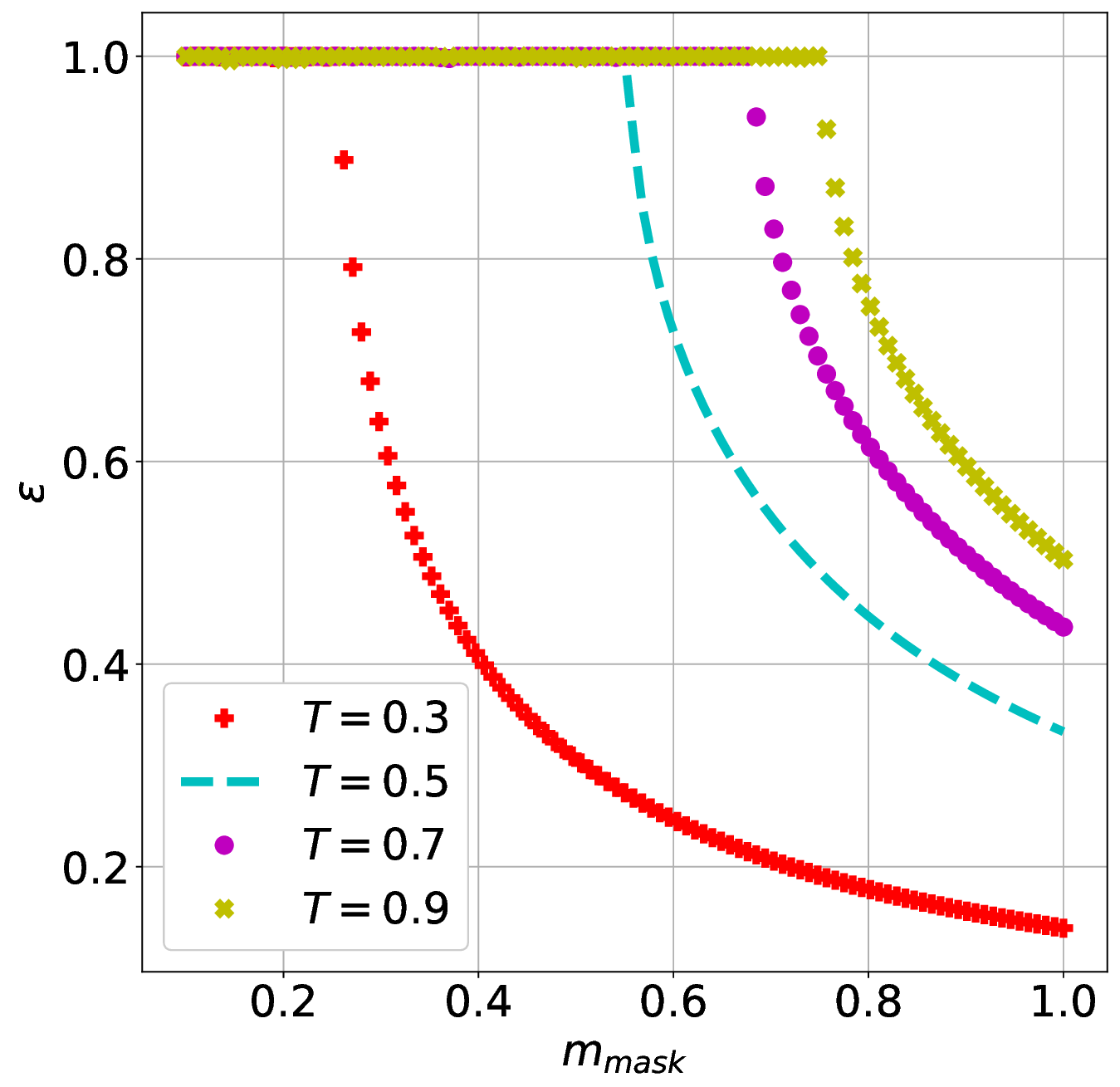}}

\caption{\sl(a) 
Epidemic \textit{boundary} that separates the region of the parameter plane that results in $R_0 \leq 1$ (i.e., epidemics are {\em not} possible) from the region that gives $R_0>1$ (i.e., epidemics {\em are}  possible). (a) Epidemic boundary shown on the parameter plane defined by the viral transmissibility $T$ and the maximum fraction of no-mask wearers $m_{\text{no-mask}}$ for different values of $\epsilon$; and (b) Epidemic boundary shown on the parameter plane defined by the  fraction of mask-wearers $m_{\text{mask}}$ and mask efficiency parameter $\epsilon$ for different values of $T$. 
In both (a) and (b), we assume two types of nodes: nodes wearing masks with the same inward and outward efficiencies $\epsilon$ and nodes wearing no masks.
$\text{Mean degree}=10$ for (a), and
$\text{mean degree}=4.5$ for (b). 
 The north and east of each curve specify the region for which epidemics are possible, while the south and west parts of each curve stand for the region where epidemics can \textit{not} occur.  
}
\label{fig:EB}
\end{figure}

In Figure \ref{fig:EB}, we present the boundary of $T - m_{\text{no-mask}}$ plane (Figure 
\ref{fig:figure2a}) and $ m_{\text{mask}} - \epsilon$ plane (Figure \ref{fig:figure2b}) that identify the epidemic threshold $R_0 = 1$ using Equation (\ref{eq:R0_epsilon}).
Here we assume two types of nodes: mask and no mask. 
$m_{\text{mask}}$ is the fraction of mask-wearers and $m_{\text{no-mask}} = 1 - m_{\text{mask}}$ is the fraction of no-mask nodes.
For simplicity of the presentation, we further set $\epsilon = \epsilon_{out} = \epsilon_{in} \in [0, 1]$ for the mask.
In Figure \ref{fig:figure2a}, for each $\epsilon$, the curves  separate the areas where epidemics can take place (north-east of the curves) from the areas where they can not (south-west of the curves).  
It is observed that with the same $T$, increasing $\epsilon$ increases the maximum $m_{\text{no-mask}}$ that is allowed in the population. 
Figure \ref{fig:figure2a} presents the boundary of $m_{\text{mask}} - \epsilon$ plane separating $R_0 = 1$.  
Similarly, in Figure \ref{fig:figure2b} for each $T$, the curves separate the areas where epidemics can take place (north-east of the curves) from the areas where they can not (south-west of the curves).  
We can see that with the same $m_{\text{mask}}$, increasing $T$  increases the minimum $\epsilon$ that is needed to prevent epidemics. 

\subsection{Expected epidemic size}
\label{subsec:es}

In this section, we compute the expected size of the epidemic -- that is, the final fraction of infected individuals of each type, conditioned on the event that the epidemic does {\em not} die out in finite time. Our method follows \cite{eletreby2020effects, gleeson2007seed, gleeson2008cascades, tian2021analysis, newman_book}. 

The analysis proceeds as follows. Let $v$ be a vertex selected uniformly at random, and recall that the local structure of the graph around $v$ is a tree with probability tending to 1 as the network size tends to infinity. The number of children of the root, $v$, follows the distribution $\{p_k \}_{k \ge 0}$ and the number of children of a later-generation vertex follows the excess degree distribution. We say that the root node is at {\it level 0}, and more generally, we say that the vertices with {\em distance} $\ell$ from the root are at {\it level $\ell$}. The epidemic is initialized by specifying the infection status of vertices far away from $v$. Formally, we specify a large positive integer $n$ as well as $\boldsymbol{\theta} : = \{\theta_i \}_{i = 1}^M$, which is a collection of values between 0 and 1. For each type-$i$ vertex in level $n$, we assume it is infected with probability $1 - \theta_{i}$, and that it is not infected with probability $\theta_{i}$. Given this initial configuration, we denote for $1 \le i \le M$ and $0 \le \ell \le n-1$ the quantity $q_{\ell, i}^{(n)}$ to be the probability that a type-$i$ vertex in level $\ell$ is {\it not} infected, given the initial configuration $\boldsymbol{\theta}$ at level $n$. The probability $q_{\ell,i}^{(n)}$ can be computed in a recursive manner, which we describe next. 

Let $u$ be a given vertex of type-$i$ in level $\ell$, so that $q_{\ell, i}^{(n)}$ is the probability that $u$ is not infected by a vertex in level $\ell + 1$. As in Section \ref{subsec:pe}, let $k_1, \ldots, k_M$ denote the number of neighbors of each type in level $\ell + 1$ and let $X_1, \ldots, X_M$ denote the number of {\it infected} neighbors of each type. Conditioned on $X_1, \ldots, X_M$, the probability that $u$ is not infected is
$$
\prod\limits_{j = 1}^M (1 - \mathbf{T}_{ji})^{X_j}.
$$
We next take an expectation over the $X_j$'s to compute the unconditional probability of non-infection. To this end, observe that the infection status of nodes in the same level are {\it independent} from each other since vertices in a given level do not have common infected descendants due to the tree-like structure of the network. Hence, since there are $k_j$ neighbors of type $j$, we have 
$$
X_j \sim \mathrm{Binomial} \left (k_j, 1-q_{\ell + 1, j}^{(n)} \right).
$$
Moreover, the $X_j$'s are independent. The probability of $u$ not being infected conditioned on $k_1, \ldots, k_M$ is therefore given by
\begin{align}
& \mathbb{E} \left [ \prod\limits_{j=1}^M (1-\mathbf{T}_{ji})^{X_j} \mid k_1, \ldots, k_M \right] \nonumber \\
& \hspace{1cm} = \prod\limits_{j = 1}^M \left( q_{\ell + 1, j}^{(n)} + \left(1 - q_{\ell + 1, j}^{(n)} \right) (1-\mathbf{T}_{ji}) \right)^{k_j} \nonumber \\
\label{eq:es_pgf}
& \hspace{1cm} =  \prod\limits_{j=1}^M \left (1 - \mathbf{T}_{ji} + q_{\ell + 1, j}^{(n)} \mathbf{T}_{ji} \right )^{k_j}.
\end{align}
Note that \eqref{eq:es_pgf} is quite similar to the PGF derived in \eqref{eq:pe_basic_pgf}, with $T_{ij}$ replaced with $T_{ji}$ and $s_j$ replaced with $q_{\ell + 1, j}^{(n)}$. Following the same steps as in Section \ref{subsec:pe}, we therefore arrive at the following recursion for $\ell \ge 1$:
\begin{equation}
\label{eq:es_later_gen}
q_{\ell, i}^{(n)} = G \left( \sum\limits_{j = 1}^M m_j \left( 1 - \mathbf{T}_{ji} + \mathbf{T}_{ji} q_{\ell + 1, j}^{(n)} \right) \right) = : F_i \left( \mathbf{q}_{\ell + 1}^{(n)} \right).
\end{equation}
For the case $\ell = 1$, we have
\begin{equation}
\label{eq:es_first_gen}
q_{0,i}^{(n)} = g \left( \sum\limits_{j = 1}^M m_j \left( 1 - \mathbf{T}_{ji} + \mathbf{T}_{ji} q_{1,j}^{(n)} \right) \right) = : f_i \left( \mathbf{q}_1^{(n)} \right).
\end{equation}
Above, $\mathbf{q}_{\ell}^{(n)} : = \{q_{\ell, i}^{(n)}\}$ is the vectorized collection of the level $\ell$ probabilities. For notational convenience, we also write $\mathbf{F} : = \{ F_i \}_{i = 1}^M$ and $\mathbf{f} : = \{ f_i \}_{i = 1}^M$. The recursions \eqref{eq:es_later_gen} and \eqref{eq:es_first_gen} imply that 
$$
\mathbf{q}_1^{(n)}  = \mathbf{F}^{(n-1)} \left( \boldsymbol{\theta} \right). $$
Conditioned on the event that the epidemic emerges, we may assume that $\theta_i < 1$ for all $1 \le i \le M$; that is, there is a positive probability that any given vertex in level $n$ is infected. Taking the limit as $n \to \infty$ shows that $\mathbf{q}_1 : = \lim_{n \to \infty} \mathbf{q}_1^{(n)}$ satisfies the fixed-point equation 
\begin{equation}
\mathbf{q}_1 = \mathbf{F}(\mathbf{q}_1).
\label{eq:q1}    
\end{equation}
The limiting probability that the root is not infected is then given by 
$$
\mathbf{q}_0 = \mathbf{f}(\mathbf{q}_1)
$$

Since $\mathbf{q}_0$ specifies the asymptotic probabilities of {\it non-infection}, $1 - q_{0,i}$ is the probability that a uniform random type-$i$ vertex is eventually infected. This is the same as the expected fraction of infected type-$i$ vertices once the epidemic has run its course. 
Note the convergence to the fixed point $\mathbf{q}_1$ is guaranteed. 
Moreover, ES and PE share the same phase transition point defined by $R_0=1$. 
More discussion on the convergence guarantee and phase transition can be found in Appendix Section \ref{apdix:convergence_phasetransition_ES}.

\section{\label{sec:numerical_results} Numerical Results}
We next present extensive numerical simulations that validate our theoretical analysis. 
In all experiments presented in this section, the contact network was generated via the configuration model with Poisson degree distribution and $1,000,000$ vertices. 
Additional experiments on networks with degree distribution following power-law with exponential cut-off appear in Appendix Section \ref{apdix:varying_T_PLC}. 
To generate the  plots, we took an average over $5,000$ independent trials where, in each trial, a new contact network was generated. We adopt $0.05$ as the threshold of epidemic emergence.

This section is organized as follows: 
Section \ref{subsec: varying_md} presents how the key epidemiological quantities are impacted by mean degrees. 
More results with varying baseline viral transmissibilities can be found in Appendix Section  \ref{apdix:varying_T}.
Section \ref{subsec: surgical_cloth} explores the effect of masks with different qualities and the implications for mask-wearing strategies given limited good quality masks.
Section \ref{subsec: inout_tradeoff} provides one of the most critical findings of this paper: PE and ES are not always behaving the same way. Source control is more important before the epidemic happens, while self-protection is essential if the epidemic already exists. 
Sensitivity analysis on different types of network structures is shown in Appendix Section \ref{apdix:sa_network_structures}.
Section \ref{subsec: degree-selection} provides a deeper look into the case where the mask type allocation depends on the degrees of the nodes, revealing the different roles nodes with different degrees play in the spreading processes.
 \subsection{Spreading process as a function of mean degree}
 \label{subsec: varying_md}
In this experiment, we assume there are three types of nodes in the population: surgical mask wearers (type-1), cloth mask wearers (type-2), and people who do not wear any masks (type-3). 
The vector $\boldsymbol{m} = [m_1, m_2, m_3]$ represents the proportions of the population for the three types of nodes. The inward efficiencies of the masks are represented by the vector $\boldsymbol{\epsilon}_{in} = [\epsilon_{\textrm{in},1}, \epsilon_{\textrm{in},2}, \epsilon_{\textrm{in},3}]$, while the  outward efficiencies are given by $\boldsymbol{\epsilon}_{out} = [\epsilon_{\textrm{out},1}, \epsilon_{\textrm{out},2}, \epsilon_{\textrm{out},3}]$. For our first experiment, we set $\boldsymbol{m} = [0.3, 0.6, 0.1]$, $\boldsymbol{\epsilon}_{out} = [0.8, 0.5, 0]$, and $\boldsymbol{\epsilon}_{in}=[0.7, 0.5, 0]$ based on the work \cite{eikenberry_masks}. In this parameter  setting, surgical masks have better inward and outward efficiencies than homemade cloth masks. 
According to recent work on the estimation of transmission probability of SARS-CoV-2 \cite{agrawal_probability_2021}, the maximum probability is 63.2\% at the source reduces exponentially to less than 1\% over a distance of 1.5 m. 
We adopt $T=0.6$ in our work as the baseline transmissibility.

\begin{figure}[h!]
    \centering
    \subfigure[Probability of Emergence]{
    \label{fig:md_prob}
    \includegraphics[width=0.4\textwidth]{ 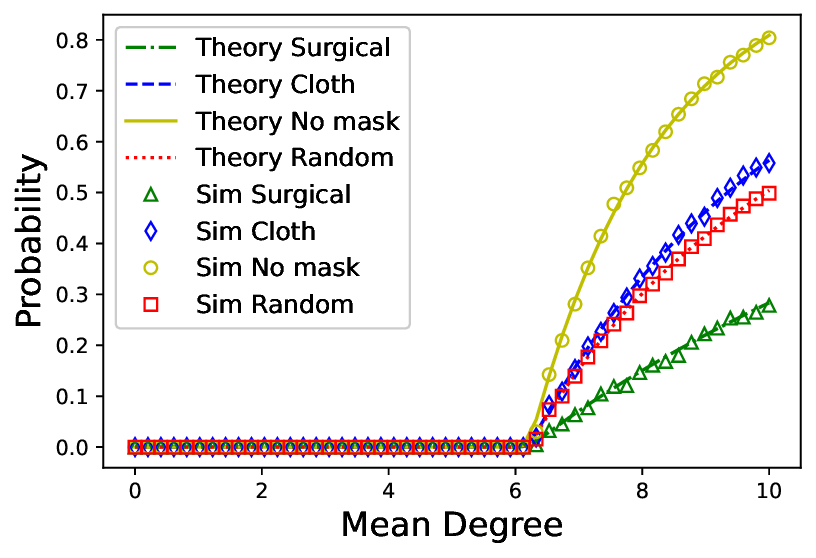}}
    \subfigure[Epidemic Size (Given Emergence)]{
    \label{fig:md_frac}
    \includegraphics[width=0.4\textwidth]{ 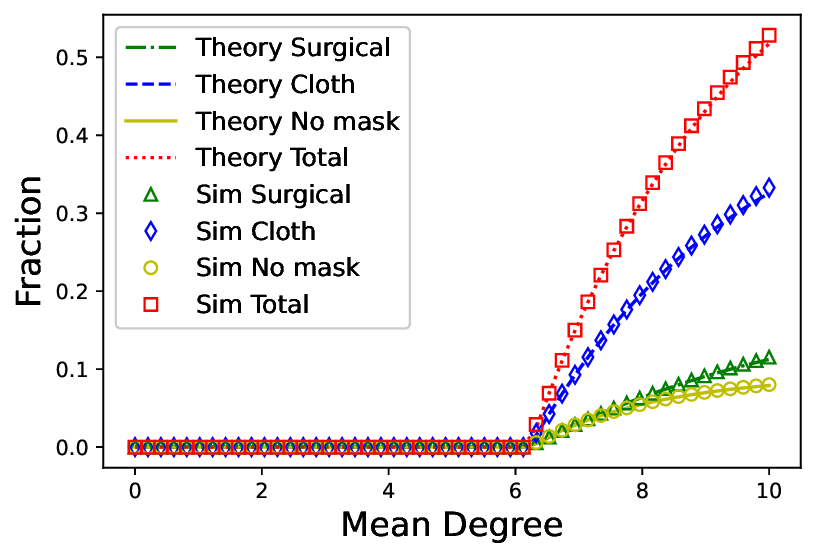}}
    \subfigure[Individual Infection Probability]{
    \label{fig:md_indiv_frac}
    \includegraphics[width=0.45\textwidth]{ 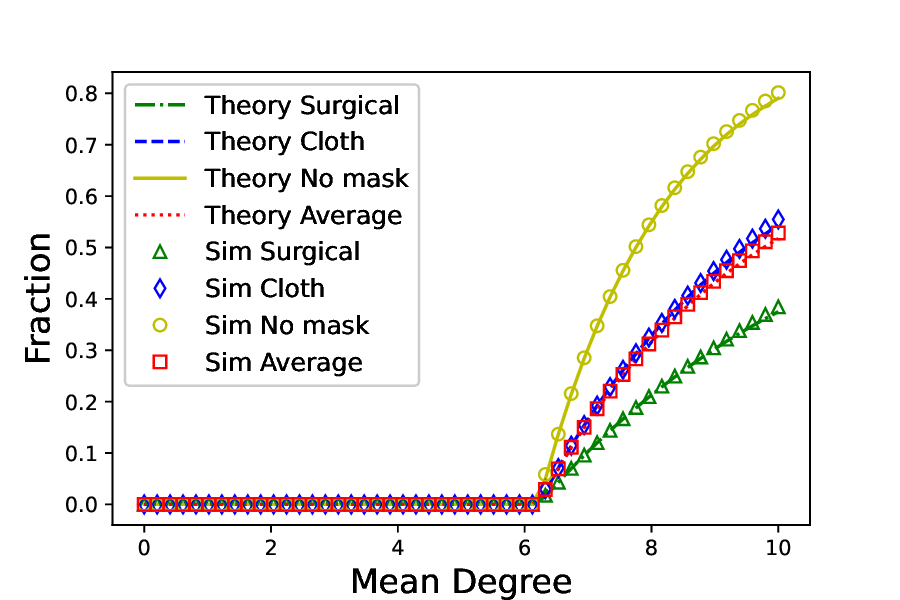}}
    \caption{\sl 
Probability of the Emergence (a), Epidemic Size (given emergence) (b), and Individual Infection Probability (c) for three mask types: surgical (green), cloth (blue), and no mask (yellow) on synthetic networks generated by the configuration model with Poisson degree distribution on a varying mean. 
$\boldsymbol{m} = [0.3, 0.6, 0.1]$, $T=0.6$, $\boldsymbol{\epsilon}_{out} = [0.8, 0.5, 0]$, and $\boldsymbol{\epsilon}_{in}=[0.7, 0.5, 0]$.
Simulation results show perfect agreement with our theoretical results with $1,000,000$ nodes and $5,000$ experiments.}
    \label{fig:md_pe_es}
\end{figure}

Figure \ref{fig:md_pe_es} studies the probability of emergence and the final epidemic size conditioned on emergence with varying mean degrees from 1 to 10. 
All figures show a perfect match between the theoretical results and simulation results. 
The $R_0$ values for mean degree = 6 and 7 are 0.921 and 1.105, respectively.
Figure \ref{fig:md_prob} compares the PE when the initial spreader is wearing  a surgical mask, a cloth mask, no mask, and random (any of the three types).
We observe that different types of initiators influence PE differently. In particular,  
the PE is lowest when the initiator wears surgical masks which have better inward and outward efficiencies than cloth masks. 
On the contrary, the probability is highest when the initiator does not wear a mask.
This is expected since mask-wearing reduces the initial transmissibility of the virus from the initiator to the later propagation. 

Figure \ref{fig:md_frac} depicts the final fraction of the infected population conditioned on epidemic emergence. The total epidemic size is the summation of the three types of infection sizes (no mask, cloth mask, and surgical mask). 
As the mean degree increases, i.e., when the average number of contacts of people in the network increases, the ES tends to increase. 
This demonstrates the effectiveness of mitigation strategies such as social distancing in reducing the total size of the infected population during a pandemic. 
\afterpage{
\begin{figure*}[ht]
    \centering
    \subfigure[Mean Degree = 8]{
    \label{fig:vary_m0_md8}
    \includegraphics[width=0.75\textwidth]{ 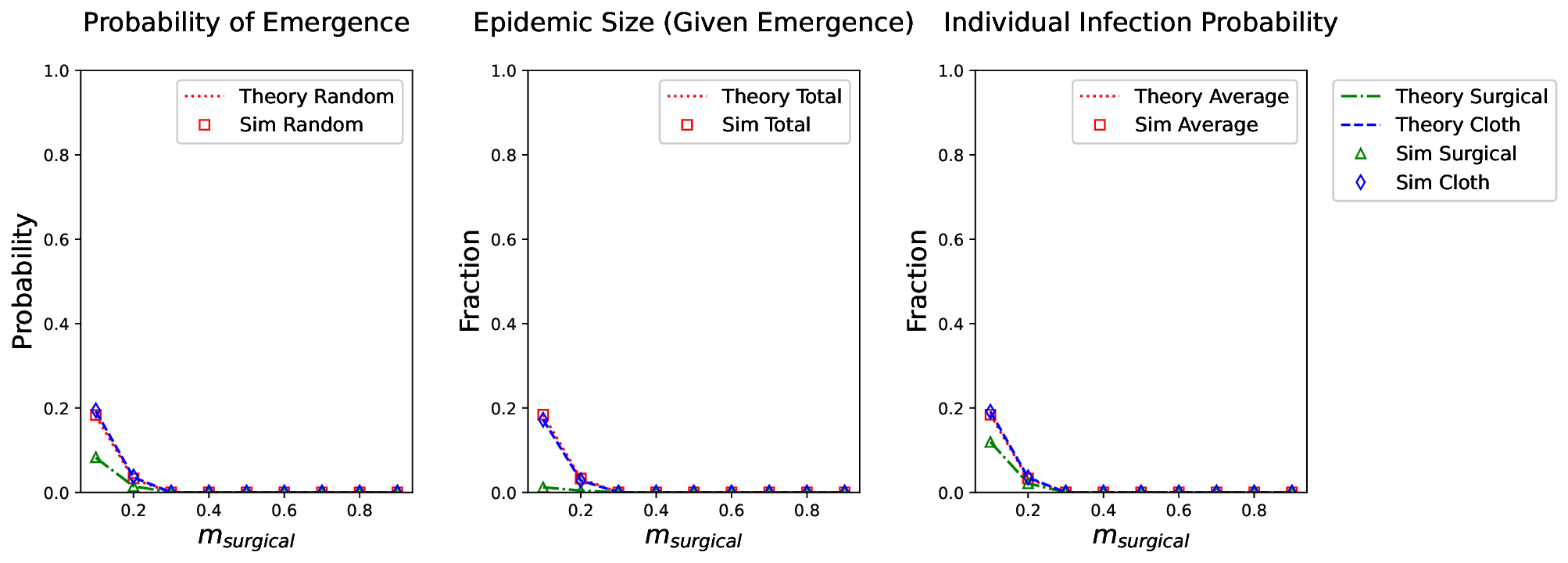}}
    \subfigure[Mean Degree = 10]{
    \label{fig:vary_m0_md10}
    \includegraphics[width=0.75\textwidth]{ 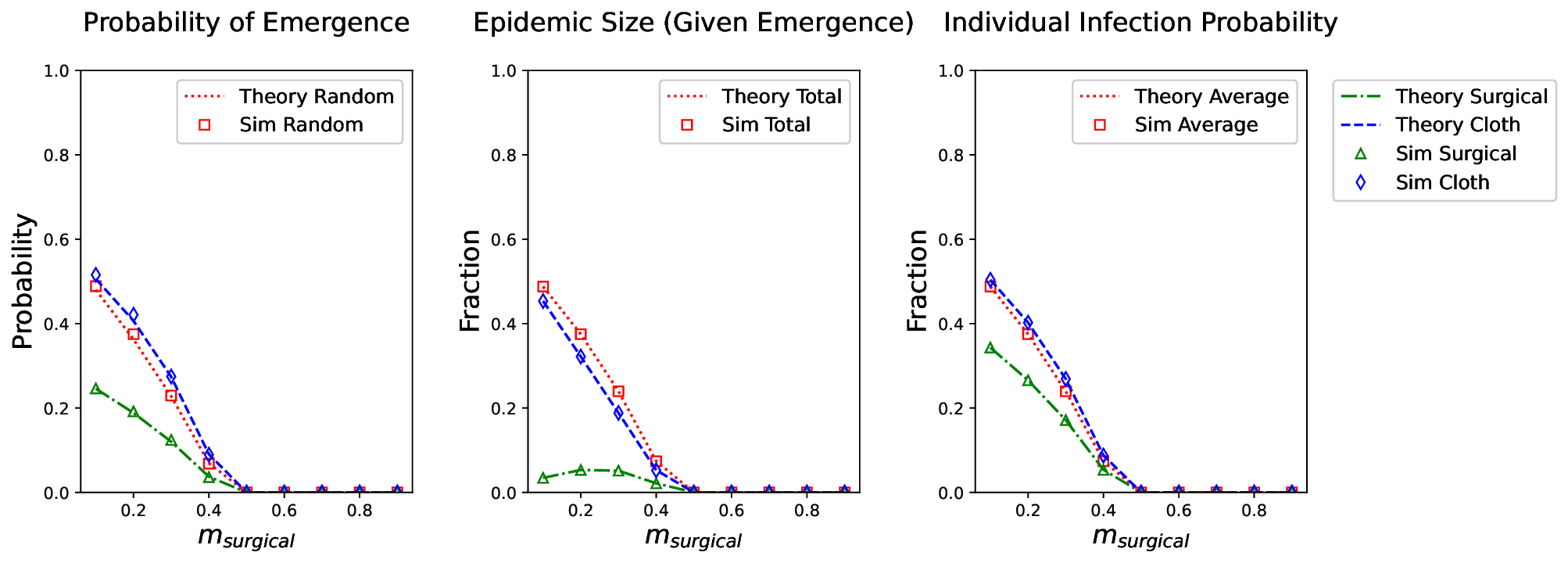}}
    \subfigure[Mean Degree = 15]{
    \label{fig:vary_m0_md15}
    \includegraphics[width=0.75\textwidth]{ 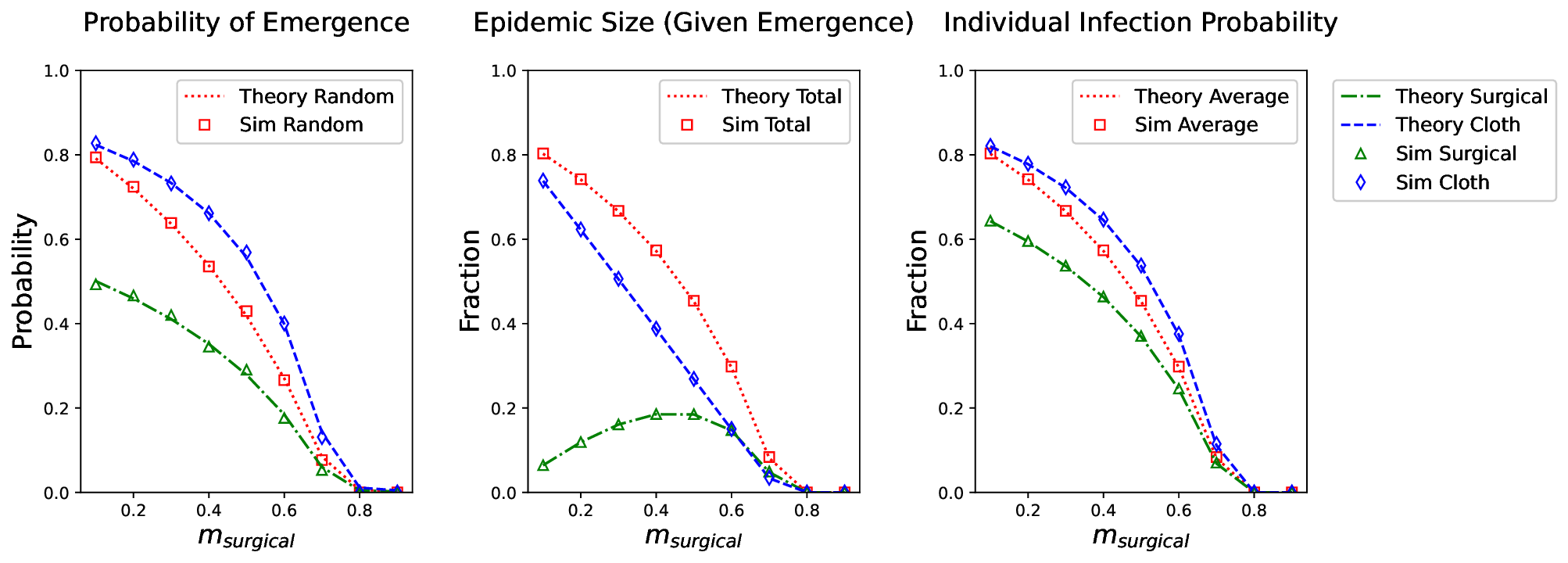}}
    \subfigure[Mean Degree = 20]{
    \label{fig:vary_m0_md20}
    \includegraphics[width=0.75\textwidth]{ 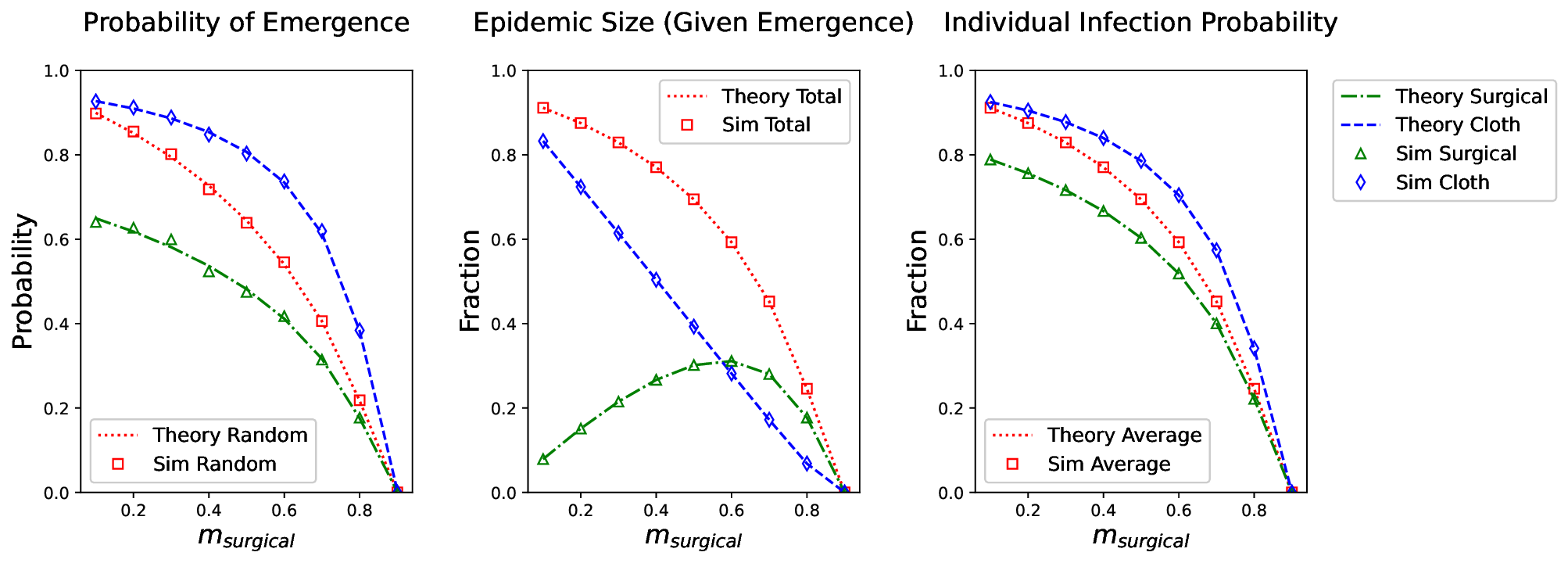}}
    \caption{\sl Probability of Emergence, Epidemic Size (given emergence), and Individual Infection Probability with an increasing proportion of surgical mask wearers ($m_{surgical}$) in the population under four mean degrees (a) 8, (b) 10, (c) 15, (d) 20 of the contact networks. $\boldsymbol{\epsilon}_{out} = [0.8, 0.5]$, and $\boldsymbol{\epsilon}_{in}=[0.7, 0.5]$ for surgical and cloth mask. $T$ = 0.6. Individual infection probability is calculated as the epidemic size of each type of mask divided by the corresponding proportion.
    Results show that increasing the proportion of surgical mask wearers can effectively compress the virus spreading. In the meanwhile, the increase in the mean degree brings down the virus-compression effect of the same proportion of surgical masks.
   The simulation is done with $1,000,000 $ nodes and $5,000$ experiments.}
    \label{fig:varing_m0_es_pe}
\end{figure*}
}

Figure \ref{fig:md_indiv_frac} presents the individual infection probability, i.e., the epidemic size of a type divided by the percentage of the same type in the population. The difference between epidemic size and individual infected probability is that the former indicates the fraction of people who are infected and of a certain type, and the latter shows the fraction of infected people within a certain type. 
The epidemic size provides us insights from a global perspective that how the infected population distributes over all types of masks, while individual infection probability gives us a view into each individual type of node. 
It is shown that no-mask wearers suffer from the largest probability of infection, followed by cloth masks and surgical mask wearers. 
No-mask wearers also have the highest increasing rate as the mean degree of the contact network increases. 
This trend also conforms with the trend of the probability of emergence shown in Figure \ref{fig:md_prob}. These results demonstrate the increased risks of infection for people wearing an inferior mask, or no mask at all.

\subsection{Comparing the effectiveness of  mask types}
\label{subsec: surgical_cloth}

We now leverage our results to examine the effect of masks with different qualities. 
For example, an interesting question to ask is: 
What should be the mask-wearing strategies to mitigate the epidemic most efficiently?

We follow the inward and outward probabilities of surgical masks and homemade cloth masks suggested by \cite{eikenberry_masks}: $\boldsymbol{\epsilon}_{out} = [0.8, 0.5]$, and $\boldsymbol{\epsilon}_{in}=[0.7, 0.5]$.
We set $T=0.6$ as it is in Section \ref{subsec: varying_md}.
Under this setting, surgical masks have better inward and outward efficiencies than cloth masks, which clearly separate the \textit{good} and \textit{bad} masks. In the later section, we will discuss the individual impact of inward and outward efficiencies in more detail.
Assume that there are only two types of nodes in the population: cloth mask wearers and surgical mask wearers. The proportion of surgical and cloth masks are given by $m_{\textrm{surgical}}$ and $m_{\textrm{cloth}}$, respectively, where we have $m_{\textrm{surgical}} + m_{\textrm{cloth}} = 1$. Figure \ref{fig:varing_m0_es_pe} illustrates the effect of changing the proportion of surgical masks  from 0.1 to 0.9 under contact networks with four different mean degrees: 8, 10, 15, and 20. 

When the fraction of surgical mask wearers increases, the probability of an epidemic with a random seed and the total epidemic size are decreasing monotonically in all four cases. 
In Figure \ref{fig:vary_m0_md8} (where the mean degree of the contact network is 8), we see that it suffices to have  30\% of the population wearing the surgical mask to make the probability of emergence drop nearly to zero; i.e., to ensure that the spreading event is unlikely to turn into an epidemic. 
In Figures \ref{fig:vary_m0_md10}-\ref{fig:vary_m0_md20}, we see that the percentage of the population wearing surgical masks needs to be at least $50\%, 80\%$ and $90\%$,
respectively, in order to make the probability of epidemics nearly zero, when the mean degree increases to 10, 15, and 20, respectively. This shows the trade-off between people having more contacts on average and the percentage of surgical mask wearers in preventing the epidemic. In particular, we conclude
 that when people are interacting with more contacts on average, significantly more people need to wear high-quality masks in order to prevent the spreading process from turning into an epidemic.

These plots also show {\em non-monotonic} trends in the epidemic size among surgical mask wearers when the mean degree is 10, 15, and 20. 
This is due to the trade-off between the growth of $m_{\textrm{surgical}}$ and the drop in the epidemic size.
The right-most column of figure \ref{fig:varing_m0_es_pe} decouples this competition. As $m_{\textrm{surgical}}$ increases,  the individual infection probability of each mask type decreases monotonically. 
The probability of emergence (the leftmost column) shares a similar tendency with individual infection probability.
Regardless of the virus spreading phase, i.e., either before or after the epidemic emerges, given various qualities of masks, our result demonstrates that it is recommended for the entire population to wear the best quality masks as much as possible to reach the most efficient virus mitigation.

\afterpage{
\begin{figure*}[ht]
    \centering
    \subfigure[$x = 10$]{
    \label{fig:vary_m0_md10_m2_1}
    \includegraphics[width=0.9\textwidth]{ 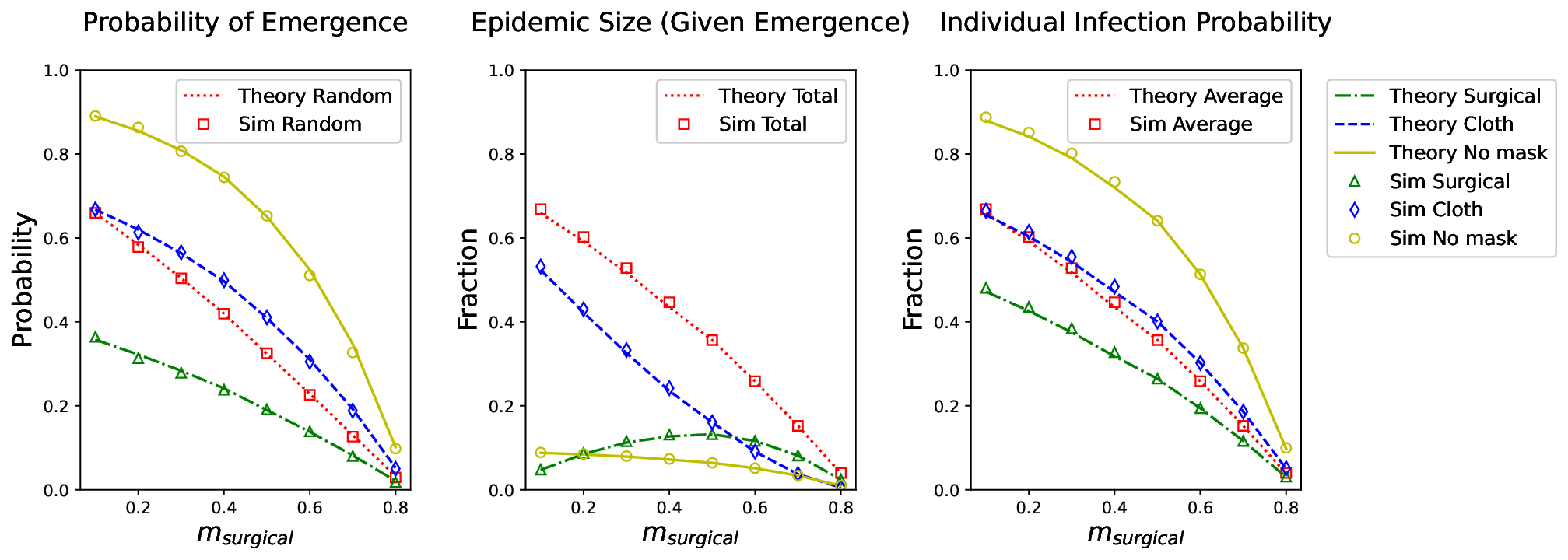}}
    \subfigure[$x = 20$]{
    \label{fig:vary_m0_md10_m2_2}
    \includegraphics[width=0.9\textwidth]{ 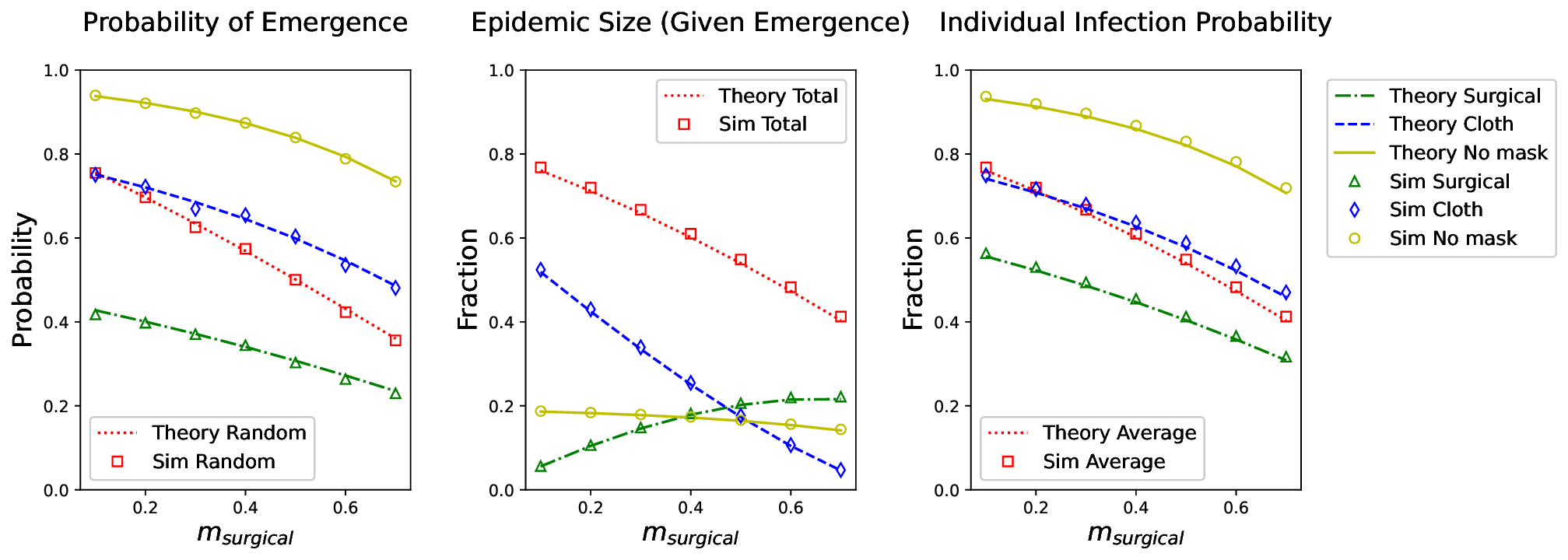}}
    \subfigure[$x = 40$]{
    \label{fig:vary_m0_md10_m2_4}
    \includegraphics[width=0.9\textwidth]{ 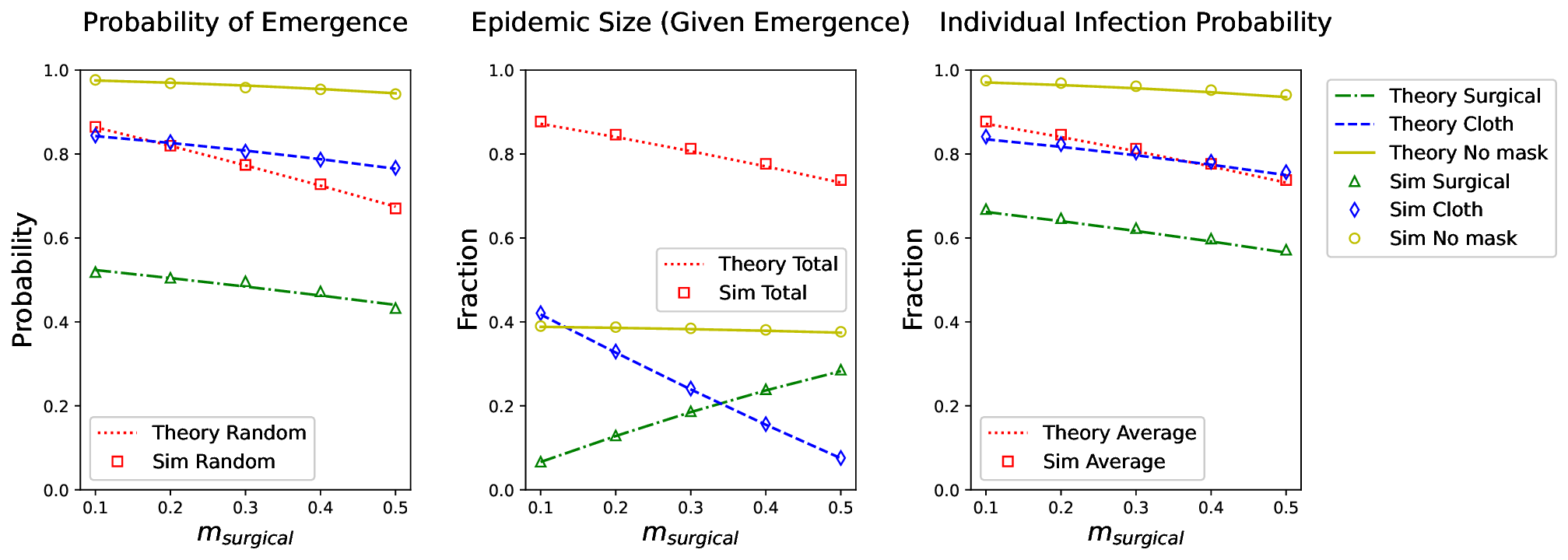}}
    \caption{
\sl Probability of Emergence, Epidemic Size (given emergence), and Individual Infection Probability with an increasing proportion of surgical mask wearers ($m_{surgical}$) in the population under a fixed mean degree = 10, for different percentages of no-mask population $x$: (a)x=10, (b)x=20, (c)x=40. $\boldsymbol{\epsilon}_{out} = [0.8, 0.5, 0]$, and $\boldsymbol{\epsilon}_{in}=[0.7, 0.5, 0]$ for surgical, cloth and no mask. $T$ = 0.6.
Compared (a) with Figure\ref{fig:vary_m0_md10}, when $x=10$, the proportion of surgical mask-wearers needs to increase from $50\%$ to $80\%$ to prevent the emergence. When $x > 10$ ((b), (c), (d)), it is not possible to prevent the emergence. In addition, the slope of the Individual Infection Probability becomes zero as $x$ increases.
The simulation is done with $1,000,000 $ nodes and $5,000$ experiments.}
    \label{fig:varing_m0_md10_es_pe}
\end{figure*}}
\afterpage{
\begin{figure*}[h!]
    \centering
    \subfigure[$x = 10$]{
    \label{fig:vary_inout_md10_m2_10}
    \includegraphics[width=0.9\textwidth]{ 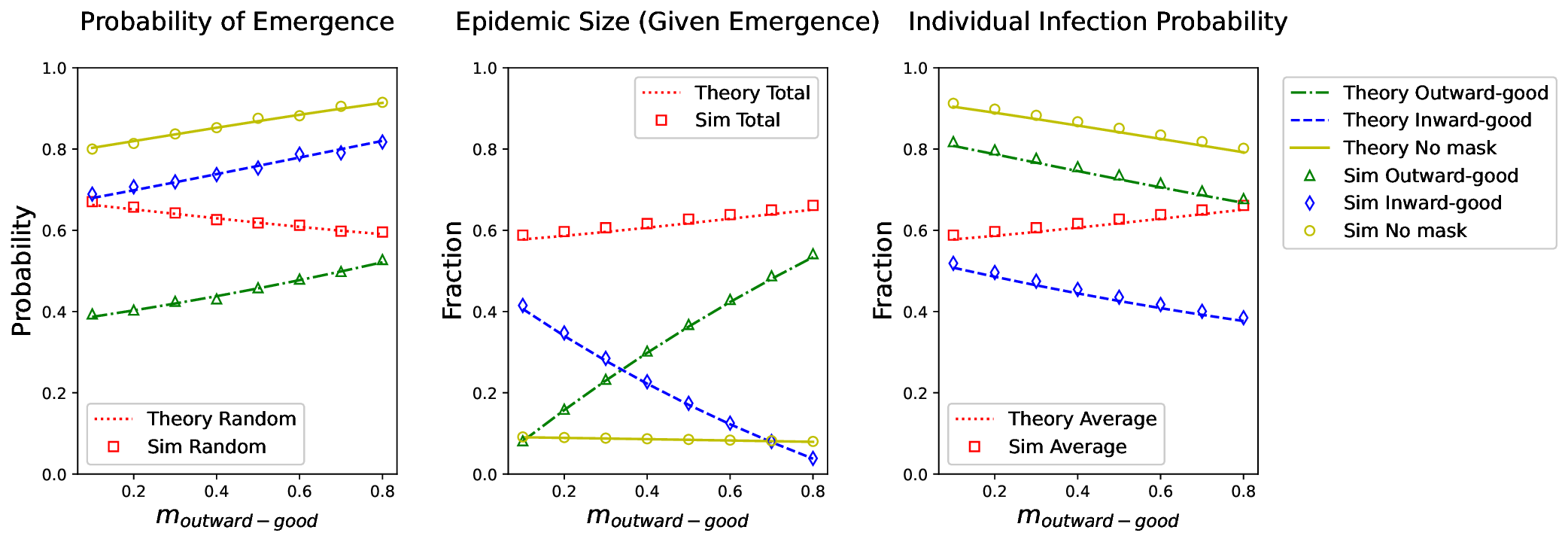}}
    \subfigure[$x = 20$]{
    \label{fig:vary_inout_md10_m2_20}
    \includegraphics[width=0.9\textwidth]{ 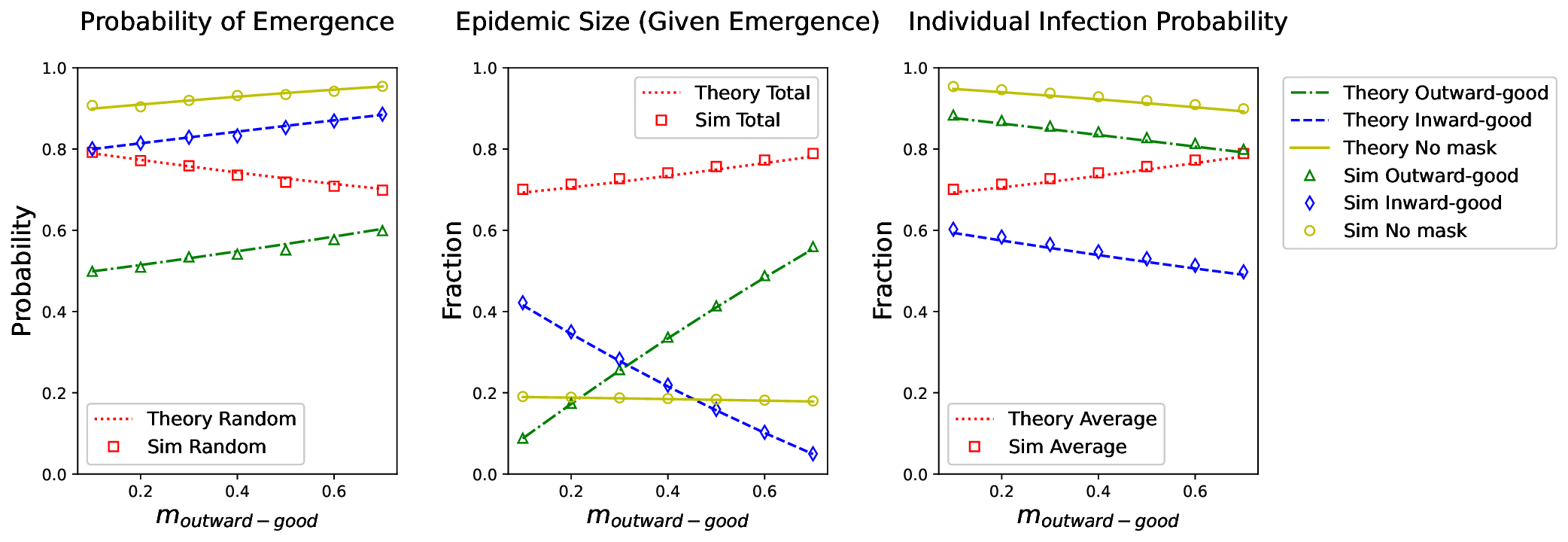}}
    \subfigure[$x = 40$]{
    \label{fig:vary_inout_md10_m2_40}
    \includegraphics[width=0.9\textwidth]{ 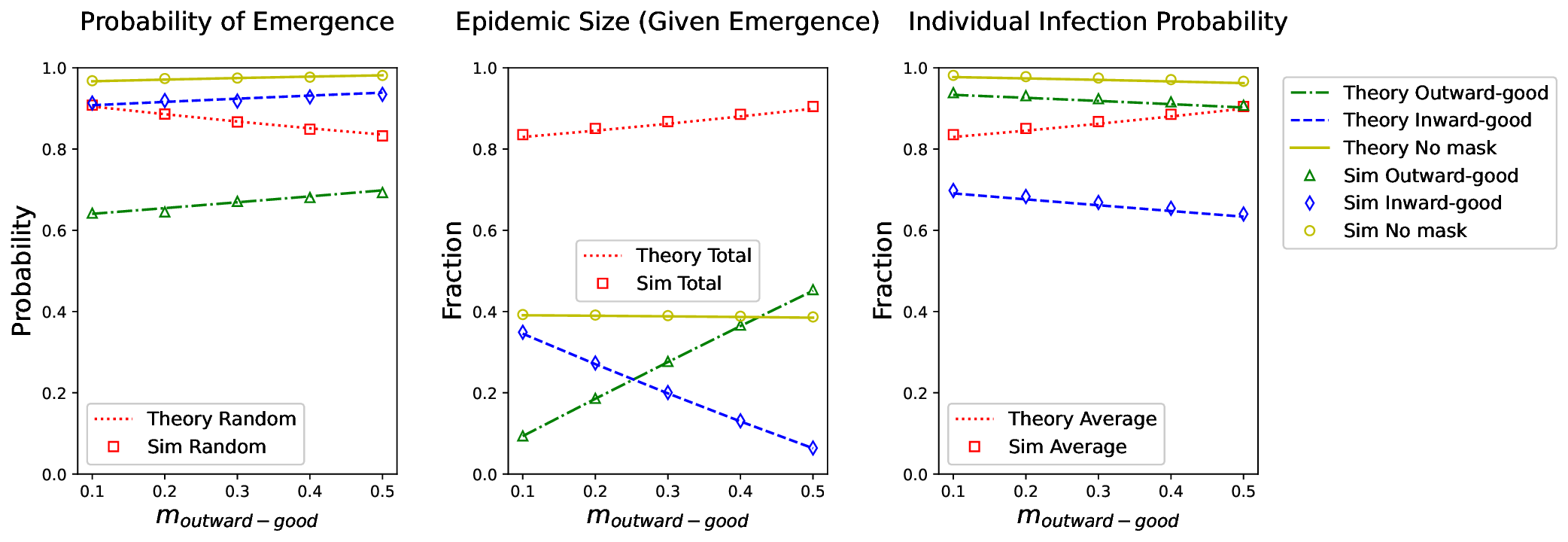}}
    \caption{\sl
    Probability of Emergence, Epidemic Size (given emergence), and Individual Infection Probability with an increasing proportion of outward-good mask wearers in the population under a fixed mean degree = 10, for different percentages of no-mask population $x$: (a)x=10, (b)x=20, (c)x=40. 
    $\boldsymbol{\epsilon}_{out} = [0.7, 0.3, 0]$, and $\boldsymbol{\epsilon}_{in}=[0.3, 0.7, 0]$ for inward-good, outward-good and no mask. $T$ = 0.6.
    As $m_{\text{outward-good}}$ increases, the probability of emergence with random seed (red) is decreasing whereas the total epidemic size given emergence (red) is increasing. This shows the two different stages these two metrics represent, where outward-good masks are less helpful before the emergence than after it.
    The simulation is done with $1,000,000 $ nodes and $5,000$ experiments.}
    \label{fig:vary_inout_md10_es_pe}
\end{figure*}
}
Next, we explore the impact of a fraction of the population not wearing any mask in our results.
In Figure \ref{fig:varing_m0_md10_es_pe}, we assume that $x\%$ of the population does not wear any masks, while the rest wear either surgical or cloth masks. 
In this case, $\boldsymbol{\epsilon}_{out} = [0.8, 0.5, 0]$, and $\boldsymbol{\epsilon}_{in}=[0.7, 0.5, 0]$ for surgical, cloth and no mask.
In other words, we set $\boldsymbol{m} = [m_{\textrm{surgical}}, m_{\textrm{cloth}}, m_{\textrm{no mask}}]$, where $m_{\textrm{no mask}} = x/100, m_{\textrm{cloth}} + m_{\textrm{no mask}}  = 1 - x/100.$
We fix the network's mean degree to 10 and generate the theoretical prediction and simulation results for the probability of emergence and epidemic size with $x=$ 10, 20, and 40. We also plot \lq\lq individual infection probability" in the rightmost column in Figure \ref{fig:varing_m0_md10_es_pe}, which is defined as the epidemic size of a given type of nodes (conditionally on the emergence of the epidemic)  divided by the proportion of that node type. Put differently, individual infection probability quantifies the odds that an individual will eventually be infected based on their mask-wearing behavior.

When $x = 10$, i.e., if  10\% of the population does not wear any masks, we see from Figure \ref{fig:vary_m0_md10_m2_1}  that even when $80\%$ of the people wear surgical masks, the epidemics still occur with positive probability; contrast this with Figure \ref{fig:vary_m0_md10} where it was sufficient for $50\%$ of the people to wear surgical masks to have PE equals zero. This means that if there are 10\% of no-mask-wearers in the population, epidemics can still occur despite the rest of the population wearing surgical masks. 

Additionally, when $x$ increases from 10 to 40, the slope of the decreasing trend for the probability of emergence and individual infection probability becomes less steep on average. This phenomenon implies that the larger the percentage of no-mask-wearers in the population, the harder for good masks to alleviate the virus spreading.

Both observations suggest that regardless of mask quality, mask-wearing should be treated as a universal requirement for the entire population for efficient epidemic prevention.

\subsection{Trade-off between inward and outward mask efficiency}
\label{subsec: inout_tradeoff}
This section explores the trade-off between inward and outward efficiencies of the masks in use, and presents one of the most critical findings on the phases of the spreading processes.
In particular,
 inward efficiency refers to the probability of a mask blocking the pathogen from coming inside the mask, while the
 outward efficiency refers to the probability of a mask stopping the pathogen from being emitted to the outside world through the mask \cite{eletreby2020effects}.
 As discussed before, 
filtration material and seal of masks could cause divergence of the two efficiencies \cite{eikenberry_masks}.
\cite{pan2021masks} thoroughly evaluated the inward and outward efficiency of 10 types of masks and a face shield under different particle size conditions.
When the particle size is around $1 \mu m$ to $2 \mu m$, the inward efficiency for a surgical mask is 25\% to 30\% while the outward efficiency is 50\% to 75\%. 
For masks made of microfiber, when the particle size is of range 2 to 5 $\mu m$, the inward efficiency (50\% to 75\%) is constantly higher than the outward efficiency (20\% to 50\%).
Besides, a total sealed face shield will have very low and similar inward and outward efficiency (both around 10\%) when the particle size is 0.5 $\mu m$. However, the outward efficiency exceeds the inward efficiency as the particle size increases: the outward efficiency reaches 75\%, and inward efficiency sticks around 25\% when the particle size grows to 5 $\mu m$. 

In this work, we use vectors $\boldsymbol{\epsilon}_{in}$ and $\boldsymbol{\epsilon}_{out}$ to represent the inward and outward efficiencies for all types of masks. 
When the inward efficiency of a mask is better than its outward efficiency, we call them inward-good masks. Similarly, we call masks with higher outward efficiency as outward-good masks.
Inward-good masks are more effective for self-protection when the subject is immersed in the environment of virus particles than blocking the virus emitted from the infected person's respiratory system. 
Similarly, outward-good masks are better at source control than the protection of the wearer. 
One practical question to ask is: if the government is  provided with both inward-good and outward-good masks, what should be the purchasing strategy? Should the government buy all inward-good masks or outward-good masks? Or, should a more complicated strategy be adopted?

Assume that there are three types of masks among the population: inward-good mask wearers, outward-good mask wearers, and people who don't wear masks, represented as type-1, type-2, and type-3 nodes, respectively.
We have the proportion vector of three types as $\boldsymbol{m} = [m_{1}, \; m_{2}, \; m_{3}]$ where $m_1 + m_2 + m_3 = 1$. 
To study the impact of mask assignment strategy for inward-good and outward-good masks, each time we fix the proportion of no-mask-wearers at $x\%$, and vary the proportion of outward-good-mask-wearers $m_{2}$ from 0.1 to 1-$x/100$. 
Based on the work by \cite{pan2021masks}, the specific efficiency parameters of the masks are selected as $\boldsymbol{\epsilon}_{out} = [0.7, 0.3, 0]$, and $\boldsymbol{\epsilon}_{in}=[0.3, 0.7, 0]$. $T=0.6$ as it is in Section \ref{subsec: varying_md}. 
Here we would like to comment on the value selection for the efficiency parameters $\boldsymbol{\epsilon}_{out}$ and $\boldsymbol{\epsilon}_{in}$ further. 
To form a fair comparison between inward-good and outward-good masks, ideally, both types are supposed to have the same average viral droplet transmission filtration power, considering both inward protection and outward protection. In other words, one type should not be strictly \textit{better} than the other type. 
Equation (\ref{eq:R0_epsilon}) provides a direct measure of each mask's average filtration power in terms of the expected number of new infections in a population where all individuals are susceptible, given the mask's inward and outward efficiencies. Thus the parameter choice for the inward-good mask and outward-good mask should follow that $(1 - \epsilon_{\textrm{out}, o})(1 - \epsilon_{\textrm{in}, o}) = (1 - \epsilon_{\textrm{out}, i})(1 - \epsilon_{\textrm{in}, i})$,  where $\epsilon_{\textrm{out}, o}$ and  $\epsilon_{\textrm{in}, o}$ ($\epsilon_{\textrm{out}, i}$ and  $\epsilon_{\textrm{in}, i}$ resp.) represent the outward and inward efficiencies for the outward-good mask (inward-good mask resp.).

Figure \ref{fig:vary_inout_md10_es_pe} shows the results of the probability of emergence, the epidemic size given emergence, and individual infection probability when the mean degree is 10, $x=$10, 20, and 40. 
An interesting result is that different strategies work best at different stages of the virus propagation process. Unlike all the previous figures, where both probabilities with random seed and total epidemic size given emergence show a decreasing trend when $m_{surgical}$ increases, Figure \ref{fig:vary_inout_md10_es_pe} displays an opposite trend between the probability of emergence and epidemic size: the probability of emergence from a random seed is reducing monotonically while the total epidemic size is increasing, as the proportion of the outward-good mask wearers $m_2$ grows. Put differently,  outward-good masks  are more helpful in terminating the spreading process before the emergence of the pandemic, whereas inward-good masks are essential to control the infection size when the pandemic already exists.

To check the generalizability of our conclusion on the trade-off between outward-good and inward-good masks, we conduct sensitivity analysis on the network structures in Appendix Section \ref{apdix:sa_network_structures}. 
We modified the network structure provided by the configuration model in four different experiment settings: 
i) increase the network's mean degree to provide a less skewed degree distribution and higher connectivity;
ii) directly remove the network structure by replacing the configuration model with a fully connected network;
iii) replace the original configuration model with \textit{configuration model with clustering }\cite{newman_random_2009} to approximate social networks with higher clustering coefficient; 
iv) replace the random graph model with a real-world dataset for studying disease spreading.
The results show that even with various modifications on the network structures, our conclusion on the trade-off between the inward and outward efficiencies remains the same. 
This demonstrates that
the configuration model provides a mathematical tractable, intuitive, and generalizable  starting point to model contact networks for studying disease-spreading processes.

We believe that this result has implications that go beyond the impact of masks and can be applied to other pandemic mitigation strategies, including prioritization of vaccines, social distancing measures, and other non-pharmaceutical interventions. Generally, it is seen that at the early stages of the virus spreading, i.e., when the infection fraction has not reached a significant percentage, a source-control-oriented strategy is crucial to prevent the epidemic from emerging. 
On the other hand, if an epidemic has already emerged and a significant fraction of the population has already been infected, it becomes most effective to implement a self-protection-oriented strategy to reduce the final fraction of the infected population. It is thus of utmost importance to develop pandemic mitigation strategies with the two distinct stages in mind, with each stage potentially having a different optimum strategy.

\subsection{Degree-based selection of inward-good and outward-good masks}
\label{subsec: degree-selection}
In the previous discussion,  we have  allocated different types of  masks to the population {\em randomly}, i.e., without any dependence on their degrees, etc.  In this section, we seek to understand whether giving high-degree nodes (e.g., people in cities) and low-degree nodes (e.g., people in villages) different types of masks would be more effective in reducing the probability and size of the epidemics compared to randomly allocating the masks.  

In particular, we assume that $x\%$ of the nodes in the population are wearing outward-good masks while the rest are wearing inward-good masks. With nodes ranked according to their degrees from highest to lowest, we consider two different mask allocation strategies as follows. Strategy 1: top $x\%$ of the nodes (with the highest degree) wearing outward-good masks; and strategy 2: bottom $x\%$ of the nodes (with the lowest degree) wearing outward-good masks. For comparison with the previous discussion, we also consider the case where the $x\%$ of the nodes wearing outward-good masks are selected uniformly at random from the entire set of nodes. 
Then, we obtain results for the probability of emergence and expected epidemic size as $x$ varies from 0 to 100. 

\begin{figure}[h]
    \centering
    \subfigure[Probability of Emergence with degree selection]{
    \label{fig:degree_selection_pe}
    \includegraphics[width=0.4\textwidth]{ 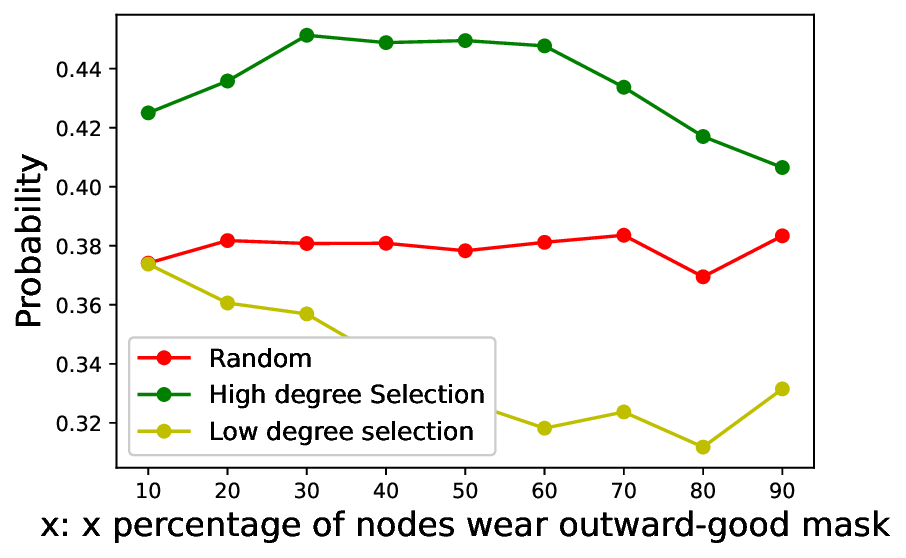}}
    \subfigure[Epidemic Size with degree selection]{
    \label{fig:degree_selection_es}
    \includegraphics[width=0.4\textwidth]{ 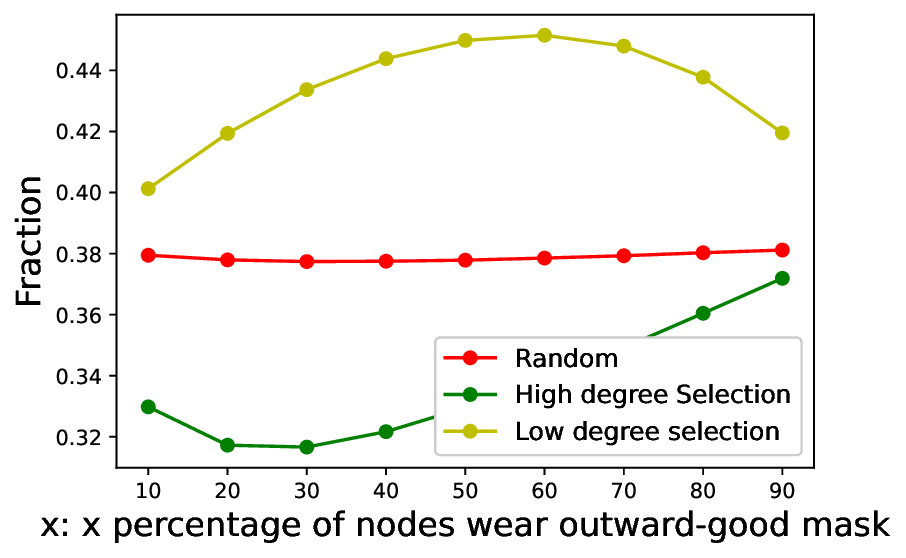}}
    \caption{\sl 
Probability of Emergence (a) and Epidemic Size (given emergence) (b) with degree selection of outward-good masks. Suppose only two mask types exist in the population: outward-good and inward-good. Three outward-good masks allocation schemes are considered: randomly allocate x percent of nodes in the population to wear outward-good masks (red); Allocate top x percent of high-degree nodes with outward-good masks (green); Allocate top x percent of low-degree nodes with outward-good masks (yellow). 
$x$ is the ratio of outward-good masks. 
$\boldsymbol{\epsilon}_{out} = [0.7, 0.3]$, and $\boldsymbol{\epsilon}_{in}=[0.3, 0.7]$ for inward-good and outward-good masks. $T=0.6$ and $\text{mean degree} = 10$.
Simulation results were obtained with $1,000,000$ nodes and $5,000$ experiments.}
    \label{fig:degree_selection}
    \vspace{-3mm}
\end{figure}

The results are presented in Figure \ref{fig:degree_selection} where we see once again an intricate difference between the pre-epidemic and post-epidemic stages of the spreading process. 
As seen in Figure \ref{fig:degree_selection_pe}, compared to randomly allocating, it is better to assign outward-good masks to low-degree nodes to reduce the probability of epidemics. 
Probability reduces by around 0.06 when $x=60$ for low-degree selection (yellow curve, strategy 2).
However, when the goal is the reduce the total fraction of infected nodes {\em given} that an epidemic took place, we see from Figure \ref{fig:degree_selection_es} that it is instead better to allocate  outward-good masks to high-degree nodes. 
Epidemic size reduces by $0.06$ when $x=30$ for high-degree selection (green curve, strategy 1).
A possible explanation for this difference is as follows.  
If our mask allocation strategy (say, Strategy 1) gives outward-good masks to high-degree nodes (and inward-good masks to low-degree nodes), the probability of transmission from an infected node to a susceptible node will be the smallest from a  high-degree node to a low-degree degree but will be the largest from a  low-degree node to a high-degree degree node. The situation will be exactly the opposite if we use Strategy 2 which gives outward-good masks to low-degree nodes; i.e., the chances of transmission will be the smallest from a  low-degree node to a high-degree degree node and largest from a  high-degree node to a low-degree degree node. For reducing the probability of epidemics, we need to consider the early stages of the spreading process,  particularly the very beginning of it. Since the {\em seed} node is selected uniformly at random, its degree will follow the network's degree distribution.  We can expect that when the seed node has a {\em high} degree, the chances of it infecting one or more of its neighbors and eventually leading to an epidemic is significant irrespective of the type of mask they are wearing. If, on the other hand, the seed node has a low degree, then there is hope that the spreading process will end early without leading to an epidemic. With this intuition, it can be seen that reducing the transmission probability from low-degree nodes to high-degree nodes would be most effective in reducing the probability of epidemics, which is achieved by Strategy 2 (i.e., by giving outward-good masks to low-degree nodes). Put differently, giving outward-good masks to high-degree nodes (which is the case for Strategy 1) leads to a situation where an infected low-degree node has a high chance of transmitting the virus to a high-degree susceptible node, increasing the probability of a single node  initiating a spreading process that leads to an epidemic. This intuition is confirmed in Figure \ref{fig:degree_selection_pe}, where we see that Strategy 1 leads to a higher probability of epidemics than the case where masks are allocated randomly without any dependence on node degrees. 

For reducing the expected  size of an epidemic that has already taken place, we can explain the results seen in \ref{fig:degree_selection_es} similarly. In this post-epidemic stage of the spreading process, there is already a critical mass of infected individuals, and we can expect to have  little chance of preventing  a  high-degree node from being infected irrespective of the mask type they are wearing; e.g., out of the many of their neighbors, several would be expected to be infected  in the post-epidemic stage, and at least one would be likely to infect this node irrespective of the corresponding mask types. Thus, the only hope for reducing the epidemic size would be to protect low-degree nodes from being infected by high-degree virus-spreaders. As previously mentioned, the probability of transmission from a high-degree node to a low-degree node is the smallest in Strategy 1, where outward-good masks are allocated to high-degree nodes.

Summarizing, we see that at the early stages of the spreading process, it is critical to protect  high-degree nodes from infection, while after the epidemic already forms, protecting other nodes from infected high-degree nodes helps more in suppressing the extension of the virus. This result echoes our previous findings that mitigation strategies for spreading processes  should be treated with two different stages in mind by focusing on source-control before the epidemic starts and self-protection after the epidemic forms.

\section{\label{sec:conclusion} Conclusion}


In this paper, we have studied an agent-based model for the viral spread on networks called the {\it multi-type mask model} in which agents wear masks of various types, leading to heterogenous probabilities of receiving and transmitting the virus. 
In particular, we performed a theoretical analysis of three critical quantities: the probability of emergence (PE), the epidemic threshold ($R_0$), and the expected epidemic size (ES), and we validated our theoretical results by comparing them against simulations and found a near-perfect match between them.

 

We then applied the model to investigate the impact of mask-wearing in realistic settings related to the control of viral spread. First, we studied the effect of allocating superior and inferior masks (e.g., surgical and cloth masks) within the population and found, naturally, that a greater prevalence of surgical masks significantly reduces PE and ES. Interestingly, we also found that when there is a significant fraction of non-mask-wearers, increasing the fraction of superior masks among mask-wearers does not significantly reduce the probability of emergence or expected epidemic size. This highlights the importance of wearing {\it some} mask -- even a potentially low-quality one -- in mitigating the spread of a virus.


Next, we examined the trade-offs between masks that are good at preventing viral transmission {\it from} an infected neighbor (inward-good masks) and masks that are good at preventing viral transmission {\it to} susceptible neighbors (outward-good masks). Strikingly, we find that the two types of masks are good for controlling different epidemiological quantities. Specifically, outward-good masks reduce PE, making them ideal for controlling the early-stage spread of the virus. On the other hand, inward-good masks reduce the ES, and hence they are most effective in mitigating the impact of an already-large pandemic. This distinction justifies the importance of source control during the early propagation stage and self-protection when a relatively large percentage of the population is infected. 

We further investigated mask assignment based on node degree. In the analysis described above, mask assignment was assumed to be independent of node degree. 
We further analyzed the situation in which masks are selected for nodes based on their degrees. 
We found that assigning high-degree nodes with inward-good masks and low-degree nodes with outward-good masks can more effectively reduce the probability of emergence than the opposite or random allocation schemes. 
In contrast, for epidemic size, high-degree nodes wearing outward-good masks and low-degree nodes wearing inward-good masks are more helpful in suppressing the  extension of the epidemic. The finding provides insights into how to allocate inward-good and outward-good masks to crowded areas and less populated regions in different stages of viral spread. 
Interestingly, the results also highlight the distinct and changing roles that high-degree and low-degree nodes play. Before the epidemic forms, low-degree node source control is critical in preventing the epidemic from happening, by which we can remove the additional paths for a single seed to initiate the epidemic. After the epidemic already exists, high-degree nodes are more impactful on the extension of the epidemic size. In this case, giving high-degree nodes outward-good masks and self-protection of low-degree nodes become essential.

Finally, we remark that there are several avenues for future research. 
There are several ways to augment the model, for instance, by considering networks with community structure, multi-type networks, and the effect of multiple strains of the virus propagating. 
It is also of great interest to explore network structure with undefined moments, such as pure power-law distributions with heavy tails.
\section*{ACKNOWLEDGMENTS}
This work was supported in part by 
the National Science Foundation through grants RAPID-2026985, RAPID-2026982, CCF-1813637, DMS-1811724, CNS-2027908, CCF-1917819, CNS-2041952, RAPID-2142997; the Army Research Office through grants \# W911NF-20-1-0204, \# W911NF-17-1-0587, and \# W911NF-18-1-0325; the C3.ai Digital Transformation Institute; Google, LLC; James S. McDonnell Foundation 21st Century Science Initiative Collaborative Award in Understanding Dynamic and Multi-scale Systems. O.Y. also acknowledges the IBM Academic Award. 
\bibliography{reference}
\clearpage
\appendix*
\section{}
\subsection{Preliminaries for Analytical Solutions}
\label{apdix:Preliminaries}
\noindent {\it Generating functions.} The PGF of the degree distribution of an arbitrary vertex in the configuration model is given by 
$$
g(x) : = \sum\limits_{k = 0}^\infty p_k x^k,
$$
where $p_k$ is the probability that a given vertex has degree $k$. It is also interesting to study the degree distribution of a node identified by following a randomly chosen {\em edge}, e.g., to characterize the distribution of the number of {\em additional} infections that a newly infected node might lead to. 
Specifically,  let  {\it excess} degree be defined as the
\lq\lq degree minus one" of a node reached by following one end of 
 an edge selected uniformly at random. The PGF of the excess degree is given by
$$
G(x) : = \sum\limits_{k = 0}^\infty \frac{k p_k }{ \langle k \rangle} x^{k-1},
$$
where $\langle k \rangle : = \sum_{k = 0}^\infty k p_k$ is the mean degree of a node. For details on the derivation, see \cite[Chapter 13]{newman_book}.
\\
\\
\noindent {\it Local structure of the configuration model.} Graphs generated according to the configuration model are known to be {\it locally tree-like} in the following sense: the local neighborhood of a uniform random vertex converges in distribution to a random tree as the number of vertices tends to infinity \cite[Chapter 12.4]{newman_book}. The number of children of the root is sampled from the degree distribution, and the number of children of a later-generation vertex is sampled from the {\it excess} degree distribution. 

\begin{figure}[h!]
    \centering
    \subfigure[Probability]{
    \label{fig:T_prob}
    \includegraphics[width=0.45\textwidth]{ 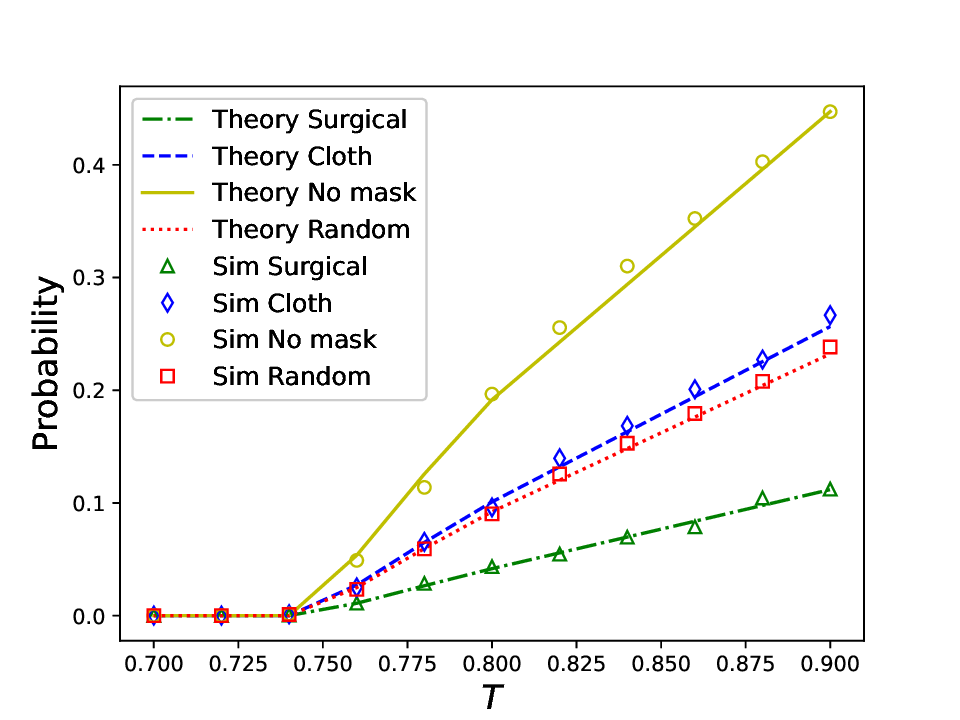}}
    \subfigure[Epidemic Size (Given Emergence)]{
    \label{fig:T_frac}
    \includegraphics[width=0.45\textwidth]{ 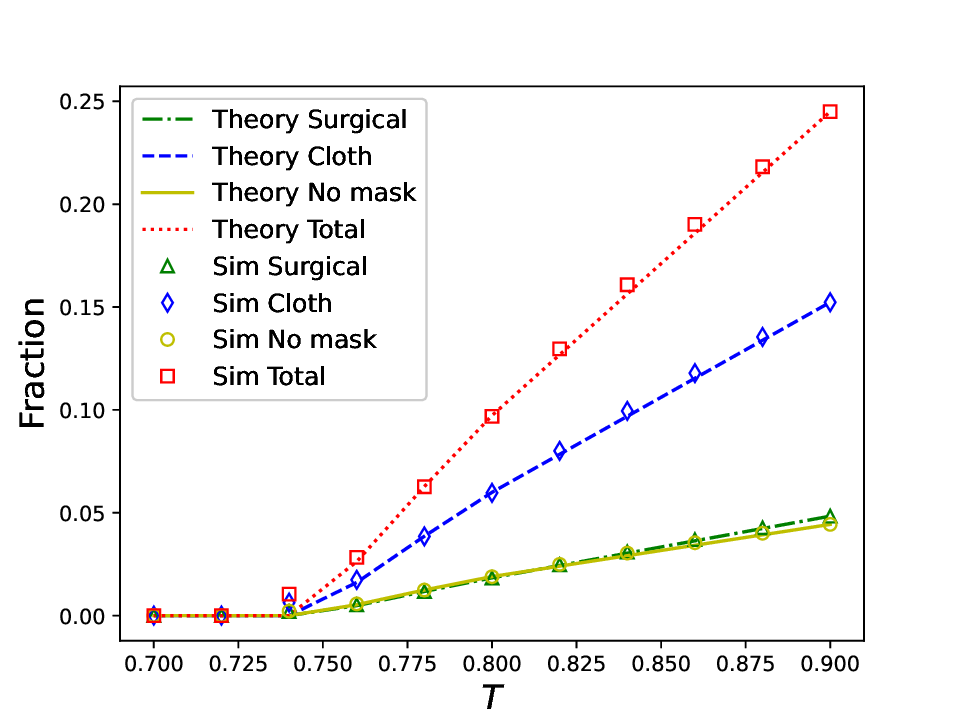}}
    \subfigure[Individual Infection Probability]{
    \label{fig:T_indiv_frac}
    \includegraphics[width=0.45\textwidth]{ 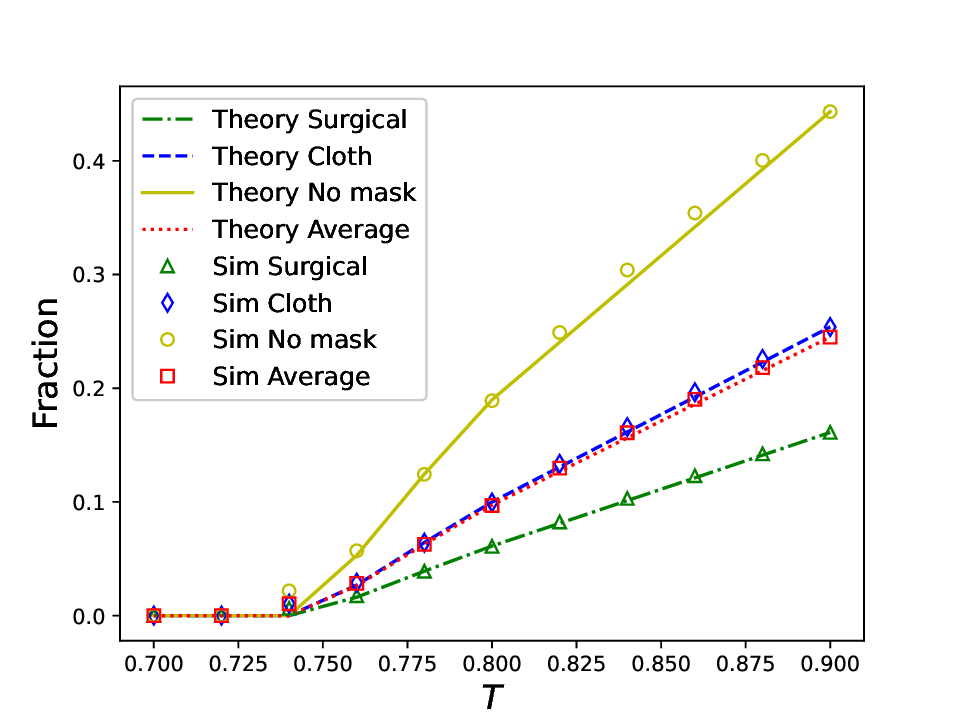}}
    \caption{\sl 
Probability of Emergence (a) and Epidemic Size (given emergence) (b) and Individual Infection Probability (c) as a function of the transmissibility of the virus ($T$). 
 Degree distribution follows $Poisson(5)$.
 $\boldsymbol{m} = [0.3, 0.6, 0.1]$.
$\boldsymbol{\epsilon}_{out} = [0.8, 0.5, 0]$, and $\boldsymbol{\epsilon}_{in}=[0.7, 0.5, 0]$ for surgical, cloth and no mask.
The simulation is done with $1,000,000 $ nodes and $5,000$ experiments.}
    \label{fig:T_pe_es}
    \vspace{-3mm}
\end{figure}

\subsection{Convergence and phase transition for ES}
\label{apdix:convergence_phasetransition_ES}
We make a few technical remarks about the computation of $\mathbf{q}_1$ in Equation (\ref{eq:q1}). Taking the limit $n \to \infty$ is a well-defined operation since the multi-type branching process corresponding to the PGF $\mathbf{F}$ is positive regular and non-singular under the assumption that $m_i > 0$ and $T_{ij} > 0$ for all $1 \le i,j \le M$ (for more details, see the discussion at the end of Section \ref{subsec:pe}). The convergence to the fixed point $\mathbf{q}_1$ is guaranteed as long as the initial condition $\mathbf{\theta}$ has positive entries \cite[Theorem V.2]{AthreyaNey}. Conditioned on the event where the epidemic emerges, this can be safely assumed. 

Just as in the PE, there is a phase transition between an ES of zero ($\mathbf{q}_0 = \mathbf{1}$) and a positive ES ($q_{0,i} < 1$ for all $1 \le i \le M$). Noting that the formulas for $\mathbf{F}$ and $\boldsymbol{\Gamma}$ are identical, except that $T_{ij}$ is replaced with $T_{ji}$, the threshold for the ES is 
\begin{equation}
\label{eq:es_threshold}
\left( \frac{\langle k^2 \rangle - \langle k \rangle }{ \langle k \rangle} \right) \rho( \mathbf{T}^\top \mathbf{m}).
\end{equation}
When \eqref{eq:es_threshold} is less than or equal to 1, $\mathbf{q}_1 = \mathbf{q}_0 = \mathbf{1}$. On the other hand, when \eqref{eq:es_threshold} is greater than 1, $q_{0,i} < 1$ for all $1 \le i \le M$. 

It can be seen that \eqref{eq:es_threshold} is exactly equal to the expression for $R_0$ defined in \eqref{eq:R0}. 

We show this by proving $\mathbf{T}^\top \mathbf{m}$ and $\mathbf{\mathbf{T}m}$ have the same spectrum, which in turn implies that $\rho(\mathbf{T}^\top \mathbf{m}) = \rho(\mathbf{\mathbf{T}m})$. This can be seen through the stronger property that the characteristic polynomials of the two matrices are identical. 
Note that for any square matrix $A$, $A$ and $A^\top$ have the same spectrum,
thus the spectrum of $\mathbf{T}^\top \mathbf{m}$ is equal to the spectrum of  $\mathbf{m}^\top\mathbf{T}$.
Besides, $\mathbf{m}$ is diagonal, so we have $\mathbf{m}^\top\mathbf{T} = \mathbf{m}\mathbf{T}$. Thus we have $\rho(\mathbf{T}^\top \mathbf{m}) = \rho(\mathbf{\mathbf{T}m})$.
The result follows from the fact that for two square matrices $A$ and $B$, $AB$ and $BA$ have the same spectrum.

Putting together the results of this section and Section \ref{subsec:R0}, we have shown that when $R_0 \le 1$, the epidemic dies out in finite time, whereas when $R_0 > 1$, the epidemic eventually infects a positive fraction of the population.

\subsection{Additional validation on analytical results}
\label{apdix:additional_validation}
\subsubsection{Spreading process as a function of viral transmissibility}
\label{apdix:varying_T}
In Figure \ref{fig:T_pe_es}, we investigate the effect of the baseline transmissibility (i.e., the transmissibility between two non-mask-wearers) on the probability of emergence and expected epidemic size. This is useful in understanding the increased risk of infection based on mask-wearing behavior in cases where high-transmissibility variants of the virus may emerge over time, e.g., the Delta variant for COVID-19. 
In Figure \ref{fig:T_pe_es}, we use the same parameter setting as in Figure \ref{fig:md_pe_es} except that the mean degree is set to 5 and $T$ varies from 0.1 to 0.9. 
As $T$ rises, the probability of emergence, epidemic size, and individual probability are all seen to increase monotonically. 
Moreover, similar to Figure \ref{fig:md_pe_es}, as the original transmissibility $T$ increases, no-mask wearers experience the highest individual infection probability as well as the highest rate of increased risk with respect to increasing baseline transmissibility. 

\subsubsection{Validation of results when the degree distribution is power-law with exponential cut-off}
\label{apdix:varying_T_PLC}
 Our analytical results for all three epidemic quantities are valid when the degree distribution is {\em well-behaved}, i.e.,  when all of its moments are finite. Put differently, our results are not restricted to networks with Poisson degree distribution.  In order to gain more insight into the consequences of the analytical results for real-world networks, we now consider a specific example of spreading processes when the contact network has power-law degree distributions with exponential cut-off. 
 Specifically, we let
\begin{equation*}
    p_k= \begin{cases}0 & \text { if } k=0 \\ \left(\operatorname{Li}_{\gamma}\left(e^{-1 / \Gamma}\right)\right)^{-1} k^{-\gamma} e^{-k / \Gamma} & \text { if } k=1,2, \ldots\end{cases}
\end{equation*}
 where $\gamma$ and $\Gamma$ are positive constants and the normalizing constant $\operatorname{Li}_{m}(z)$ is the $m$th polylogarithm of $z$; i.e.,  $\operatorname{Li}_m(z)=\sum_{k=1}^{\infty} \frac{z^k}{k^m}$.

\begin{figure}[!h]
    \centering
    \subfigure[Probability]{
    \label{fig:T_prob}
    \includegraphics[width=0.4\textwidth]{ 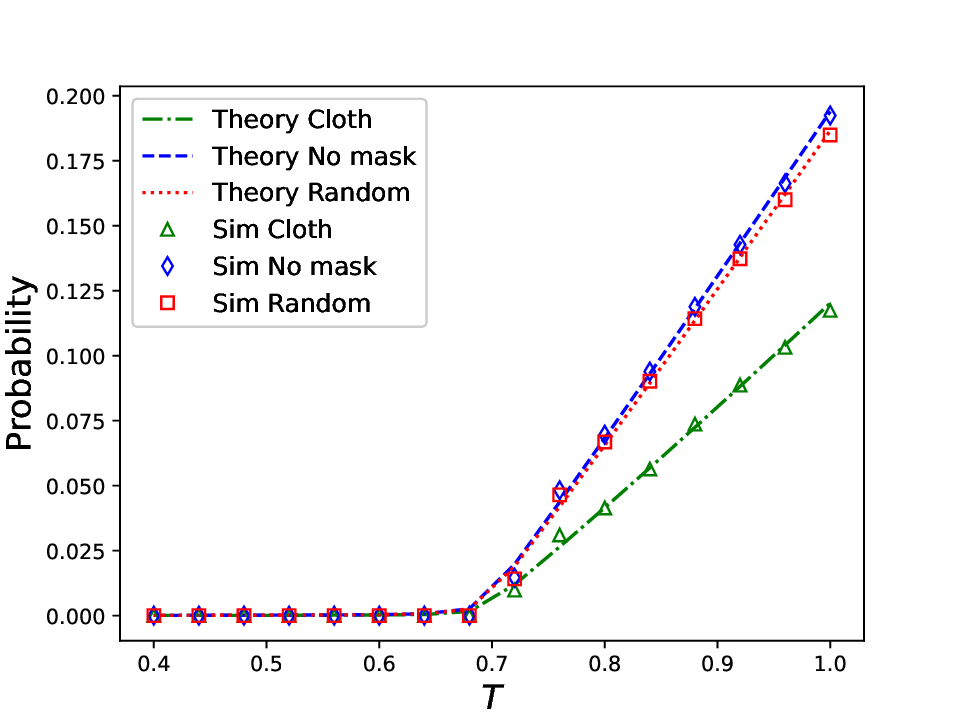}}
    \subfigure[Epidemic Size (Given Emergence)]{
    \label{fig:T_frac}
    \includegraphics[width=0.4\textwidth]{ 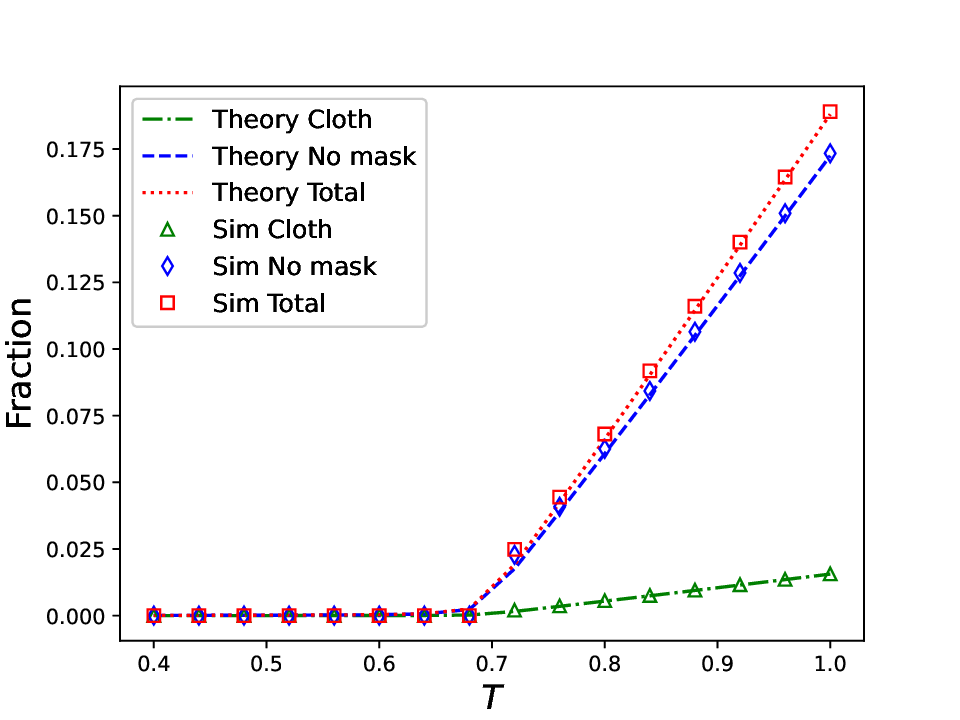}}
    \subfigure[Individual Infection Probability]{
    \label{fig:T_indiv_frac}
    \includegraphics[width=0.4\textwidth]{ 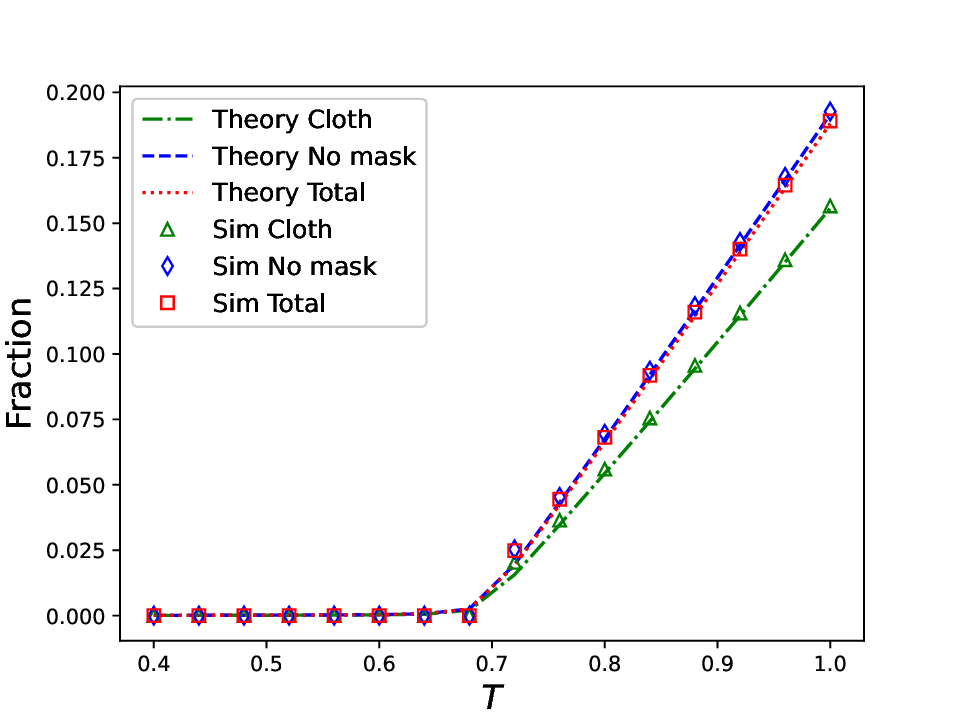}}
    \caption{\sl 
(a) Probability of Emergence,  (b) Epidemic Size (given emergence),  and (c) Individual Infection Probability as a function of the transmissibility $T$ of the virus. The underlying degree distribution is generated through power-law with exponential cut-off where the power exponent equals 2.5 and the cut-off equals 10. The parameter choices are as follows: $\text{Mean degree}=1.028$,
$\boldsymbol{m} = [0.1, 0.9]$, $\boldsymbol{\epsilon}_{out} = [0.4, 0]$, and $\boldsymbol{\epsilon}_{in}=[0.2, 0]$ for cloth and no mask, respectively.
The network consists of $1,000,000 $ nodes and we show the average results over $5,000$ experiments. 
We see that our analytical results still show a near-perfect match with simulation results when the degree distribution is power-law with exponential cut-off.}
    \label{fig:T_pe_es_PLC}
    \vspace{-3mm}
\end{figure}

 We choose power-law distributions with exponential cut-off here because they are applied to a wide range of real-world networks \cite{Newman_2002, leicht_percolation_2009}, and they are \textit{well-behaved}, i.e., have finite moments. Figure \ref{fig:T_pe_es_PLC}
 demonstrates that our theoretical results match simulations well for networks for $\gamma=2.5$ and $\Gamma=10$, where these two parameter values are selected in line with \cite{yagan2013conjoining}.
 
\subsection{Sensitivity analysis on network structure}
\label{apdix:sa_network_structures}
In section \ref{subsec: inout_tradeoff}, we discussed the trade-off between inward-good and outward-good masks: different strategies should be considered at different stages of the virus spreading process. Figure \ref{fig:vary_inout_md10_es_pe} presents the opposite trends between the probability of emergence with random seed and total epidemic size: as the percentage of outward-good masks increases, the probability of emergence decreases but the epidemic size increases. Outward-good masks  are useful for preventing the epidemic from happening, while inward-good masks are helpful in controlling the infection size if the epidemic already exists. 
In this section, we conduct sensitivity analysis to explore further the impact of the network structure on the trends exhibited by figure \ref{fig:vary_inout_md10_es_pe}. 
The analysis aims at demonstrating the robustness of our conclusion on the trade-off between outward-good and inward-good masks.
 Four experiment settings are considered:
i) repeating experiments of figure \ref{fig:vary_inout_md10_es_pe} with simple modifications as follows: increase the mean degree of the Poisson degree distribution from 10 to 30 and decrease the original virus transmissibility $T$ from 0.6 to 0.2;
ii) replacing the network generated by the configuration model with a small fully connected network with the number of nodes 1001 and mean degree 1000 and adjusting $T$ from 0.6 to 0.006. 
iii) replacing the original configuration model with \textit{configuration model with clustering} proposed by Newman in  \cite{newman_random_2009}; 
iv) replacing the generated random graph with \textit{Haslemere dataset}, a real-world public dataset on human social contacts collected specially for modeling infectious disease dynamics \cite{noauthor_sparking_nodate}. 
We name the above four settings from experiment setting 1 to experiment setting 4, respectively. 
Experiment setting 1 increases the connectivity of the network generated by the configuration model, providing a less skewed and more even degree distribution. 
Experiment setting 2 removes the topological structure explicitly by using a small fully connected network instead of the network generated by the configuration model.  We set the network size to 1000 in experiment setting 2 due to the computation complexity of fully connected networks.
Experiment setting 3 increases the clustering coefficient of the generated random graph to approximate real-world social networks. This model generalizes the standard \textit{configuration model}, which specifies the number of edges connected to each vertex. Instead, in \textit{configuration model with clustering}, both the number of single edges and the number of triangles are specified.
It incorporates clustering in a simple, sensible fashion.
Experiment setting 4 uses a real-world contact network for epidemic simulation, a part of the BBC documentary `Contagion! The BBC Four Pandemic' \cite{klepac_contagion_2018, noauthor_sparking_nodate}. 
The data is high-resolution, collected from residents of the town of Haslemere, where the first evidence of UK-acquired infection with COVID-19 was reported in late February 2020 \cite{noauthor_coronavirus_2020}. We call this dataset \textit{the Haslemere dataset} in this paper. 
In experiment 1 and 2 settings, we adjust $T$ to avoid the situation where the probability of emergence reaches 1 regardless of the percentage of outward-good masks by keeping the product of $T$ and mean degree as a constant.

\begin{figure*}[!h]
    \centering
    \subfigure[$x = 10$]{
    \label{fig:vary_inout_md30_m2_10}
    \includegraphics[width=0.9\textwidth]{ 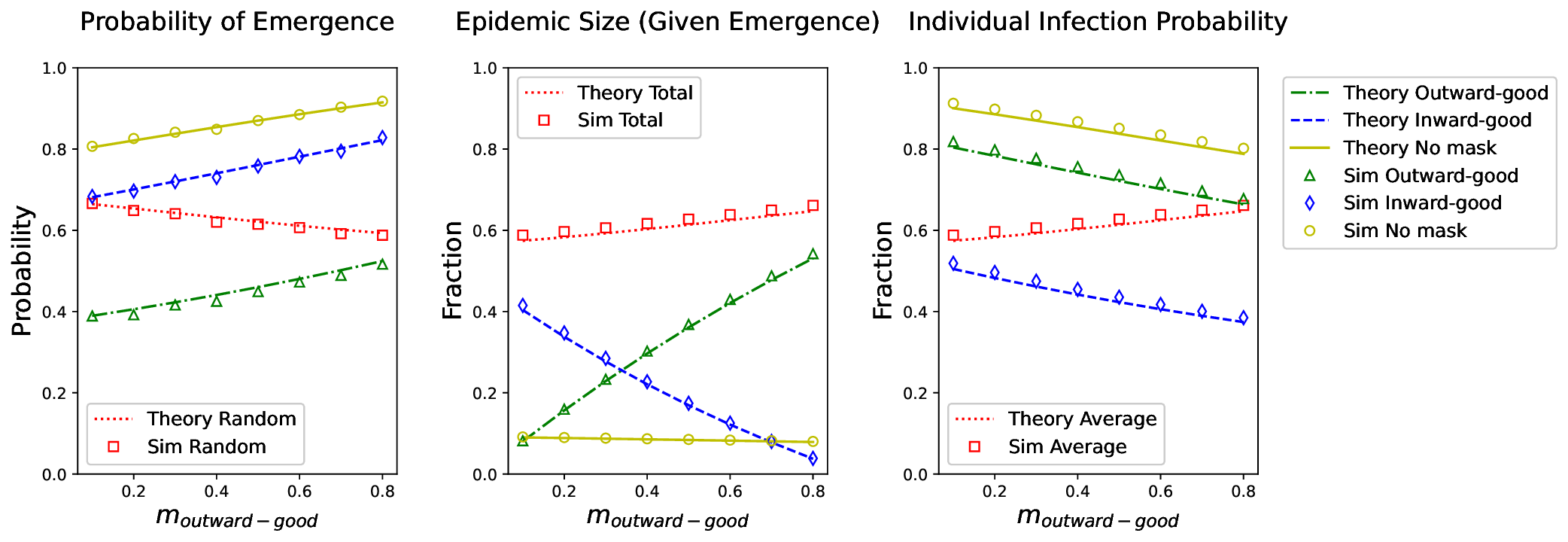}}
    \subfigure[$x = 20$]{
    \label{fig:vary_inout_md30_m2_20}
    \includegraphics[width=0.9\textwidth]{ 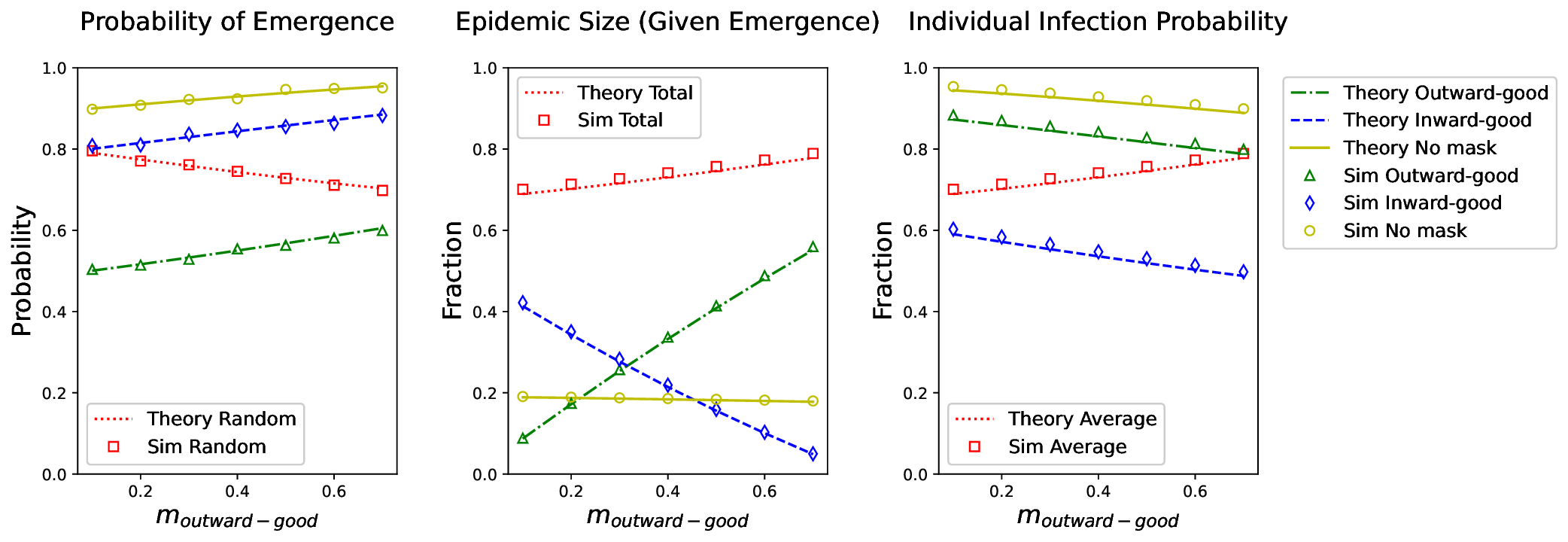}}
    \subfigure[$x = 40$]{
    \label{fig:vary_inout_md30_m2_40}
    \includegraphics[width=0.9\textwidth]{ 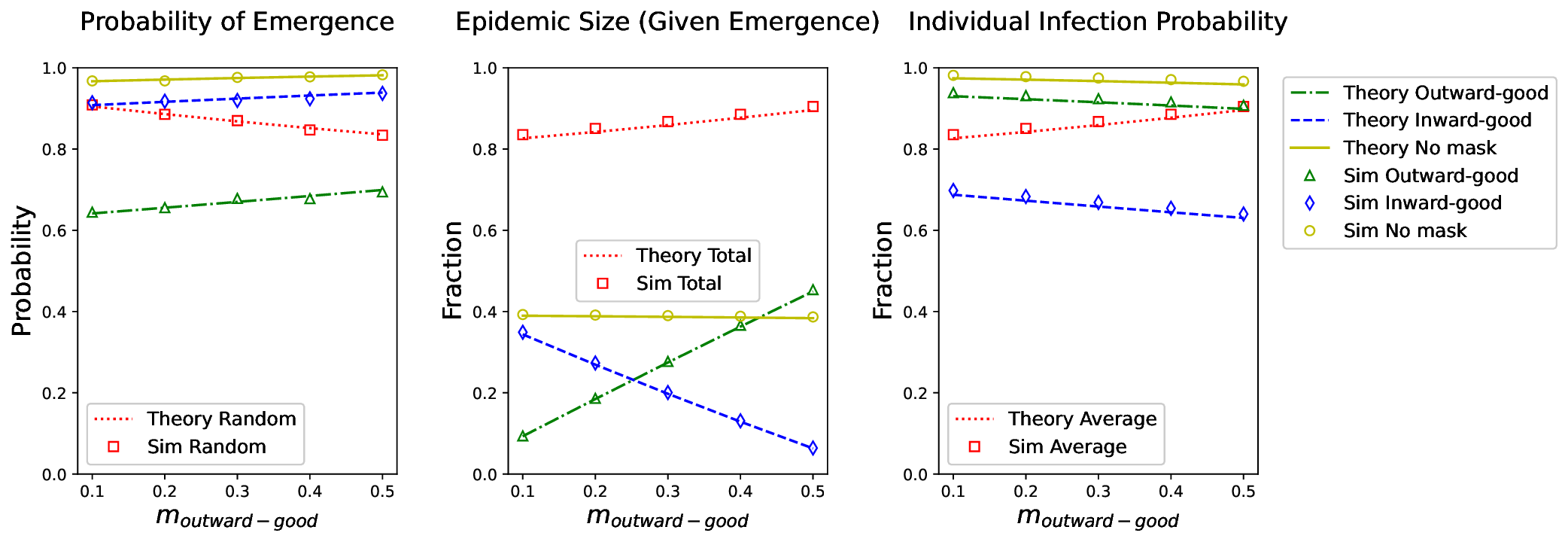}}
    \caption{\sl
    Probability of Emergence, Epidemic Size (given emergence), and Individual Infection Probability with an increasing proportion of outward-good mask wearers in the population under a fixed mean degree = 30, for different percentages of no-mask population $x$: (a)x=10, (b)x=20, (c)x=40. 
    $\boldsymbol{\epsilon}_{out} = [0.7, 0.3, 0]$, and $\boldsymbol{\epsilon}_{in}=[0.3, 0.7, 0]$ for inward-good, outward-good and no mask. $T$ = 0.2.
    Results resemble figure \ref{fig:vary_inout_md10_es_pe} after tripling the mean degree and reducing $T$. 
    The conclusion on the trade-off between inward and outward efficiency still holds.
    The simulation is done with $1,000,000 $ nodes and $5,000$ experiments.}
    \label{fig:vary_inout_md30_es_pe}
\end{figure*}

Figure \ref{fig:vary_inout_md30_es_pe} shows the result for experiment setting 1: the probability of emergence, the epidemic size given emergence, and the expected fraction of infection, for the inward-good mask, outward-good mask, and no mask. The percentage of the population who wears no mask, $x$, is set to 10, 20, 40 from figure \ref{fig:vary_inout_md30_m2_10} to figure \ref{fig:vary_inout_md30_m2_40}.  The efficiency parameters are set the same as it is in figure \ref{fig:vary_inout_md10_es_pe}: $\boldsymbol{\epsilon}_{in} = [0.7, 0.3, 1]$, and $\boldsymbol{\epsilon}_{out}=[0.3, 0.7, 1]$.
The mean degree is 30, and transmissibility $T = 0.2$.
We can see that figure \ref{fig:vary_inout_md10_es_pe} and figure \ref{fig:vary_inout_md30_es_pe} result in the same trends: the probability of emergence with random seed decreases, and the total epidemic size increases as the percentage of outward-good masks increases. 
Moreover, all three quantities: the probability of epidemics, epidemic size, and individual infection probability in figure \ref{fig:vary_inout_md30_es_pe}  are close to their counterparts in figure \ref{fig:vary_inout_md10_es_pe}.
There is no significant difference between the results shown in both figures. 
This is understandable because even though we alter the underlying network structure by tripling the number of paths on average for the virus to spread, we reduce the transmissibility at the same time. This supports the fact that the spreading process is a function of multiple factors. 
Our result incorporates the trade-off among those factors and thus displays robustness under various conditions.

\begin{figure*}[h]
    \centering
    \subfigure[$x = 10$]{
    \label{fig:vary_inout_fully_md1k_m2_10}
    \includegraphics[width=0.9\textwidth]{ 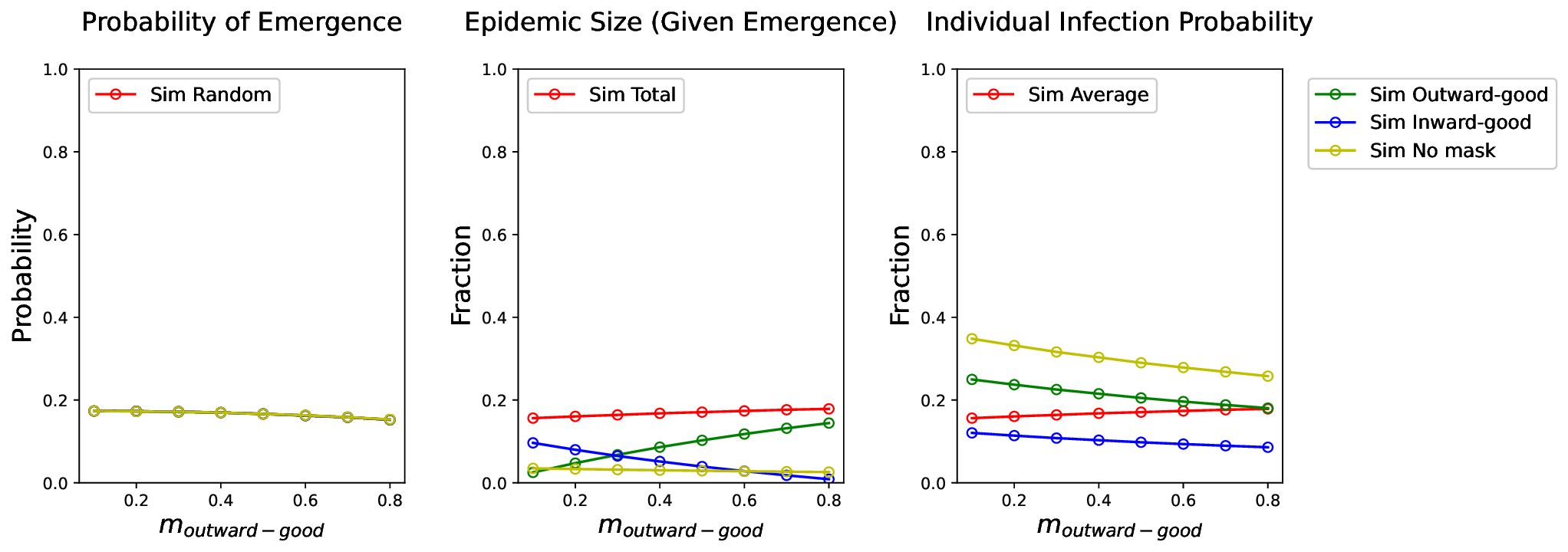}}
    \subfigure[$x = 20$]{
    \label{fig:vary_inout_fully_md1k_m2_20}
    \includegraphics[width=0.9\textwidth]{ 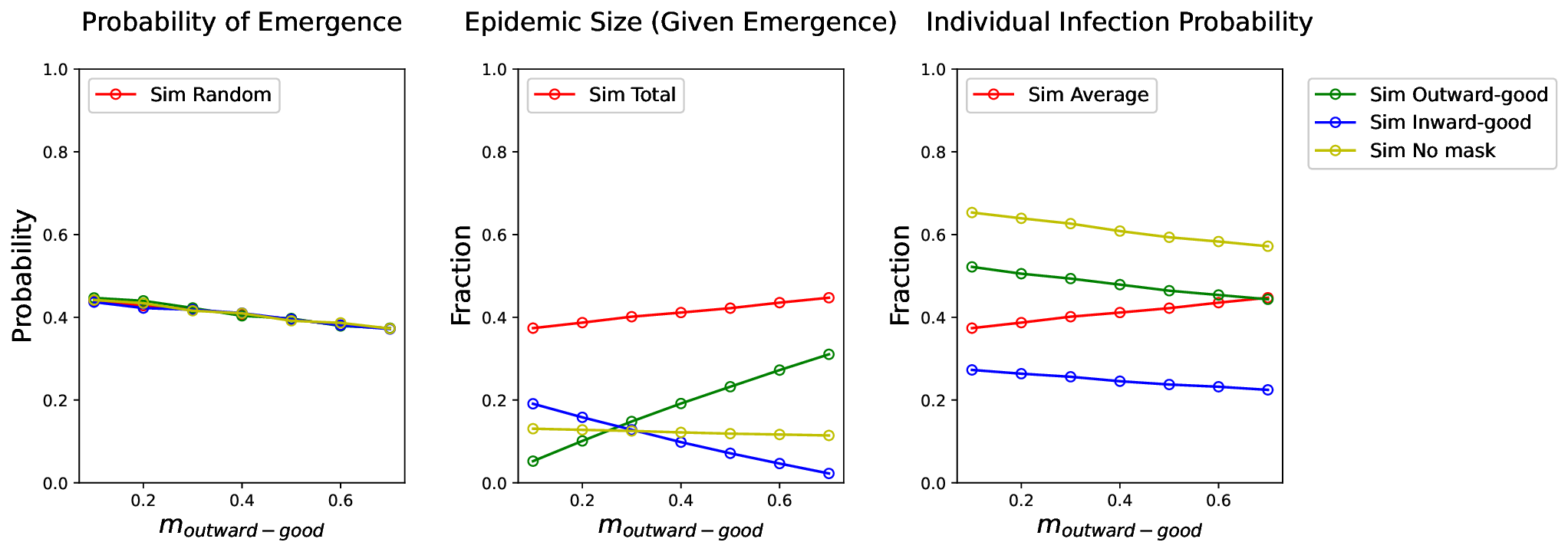}}
    \subfigure[$x = 40$]{
    \label{fig:vary_inout_fully_md1k_m2_40}
    \includegraphics[width=0.9\textwidth]{ 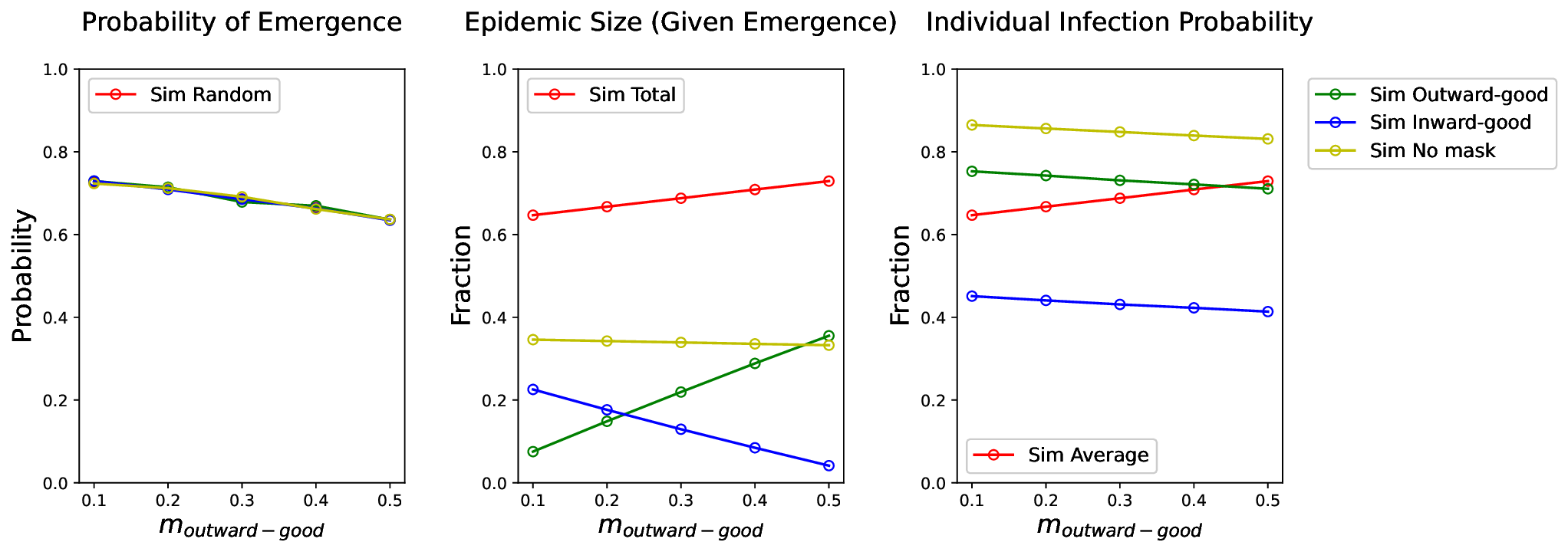}}
    \caption{\sl
    Probability of Emergence, Epidemic Size (given emergence), and Individual Infection Probability with an increasing proportion of outward-good mask wearers ($m_{outward-good}$) in the population under a fixed mean degree = 1000, for different percentages of no-mask population $x$: (a)x=10, (b)x=20, (c)x=40. 
    $\boldsymbol{\epsilon}_{out} = [0.7, 0.3, 0]$, and $\boldsymbol{\epsilon}_{in}=[0.3, 0.7, 0]$ for inward-good, outward-good and no mask. $T$ = 0.006.
    The trend that as $m_{outward-good}$ increases, the probability of emergence with random seed decreases and total epidemic size increases remains robust. 
    This supports the conclusion that we should focus on source control before epidemics and self-protection if an epidemic happens. 
    The simulation is done with $1,000,000 $ nodes and $5,000$ experiments.}
    \label{fig:vary_inout_fully_md1k_es_pe}
\end{figure*}

Figure \ref{fig:vary_inout_fully_md1k_es_pe} presents the simulation results of experiment setting 2. The contact network is a fully connected network with a node size of 1001 and a mean degree of 1000. Original virus transmissibility $T$ is 0.006. Other parameter settings are the same as experiment setting 1.
Figure \ref{fig:vary_inout_fully_md1k_es_pe} shows that the trend that the probability of emergence with random seed decreases and the total epidemic size increases as the percentage of outward-good mask wearers increases still holds. 
This again shows that, even though the result is derived under certain assumptions of network structures, however, our conclusion is generalizable. 
In figure \ref{fig:vary_inout_fully_md1k_es_pe}, one observation is noticeable: the probability of emergence for outward-good, inward-good, and no mask wearers overlap. 
This is potentially due to the fact that all the nodes have the same degree. 
Based on the observations from section \ref{subsec: degree-selection}, high-degree and low-degree nodes play different roles in the two phases of virus spreading.
Before the epidemic happens, it is critical to protect low-degree nodes by assigning them inward-good masks in order to reduce extra pathways for high-degree nodes to get infected.
However, if everyone in the network has the same degree and connects to everyone else, the increase of outward-good masks becomes important to all the nodes in the network. 
As for epidemic size, the results for all three types of users are not overlapping in figure \ref{fig:vary_inout_fully_md1k_es_pe}. 
They resemble the trends in figure \ref{fig:vary_inout_md10_es_pe} and figure \ref{fig:vary_inout_md30_es_pe}. 
This is because, at this stage of the spreading process, most nodes are infected, whether all the nodes are of the same degree or the network exhibits a wide range of degree distribution.

At this stage, the more inward-good masks (less outward-good masks) are assigned, the better self-protection capacity the population gains. This is also observed from the consistently decreasing trends of individual infection probability from figure \ref{fig:vary_inout_md10_es_pe}, \ref{fig:vary_inout_md30_es_pe} and \ref{fig:vary_inout_fully_md1k_es_pe}. 
Due to the complicated interplay of various factors such as network structure, mask allocation, and virus spreading dynamics, it comes as no surprise that the removal of the network structure features does influence the results in some sense. For example, we no longer see the discrepancy in the probability of emergence for three types of agents;  the exact values and slope of the curves are not the same, etc. Nevertheless, we observe that under most conditions, source control is critical before the epidemic happens, and self-protection is more helpful if the epidemic already exists remains.

\begin{figure*}[!h]
    \centering
    \subfigure[Probability of Emergence]{
    \label{fig:pe_ccm}
    \includegraphics[width=0.4\textwidth]{ 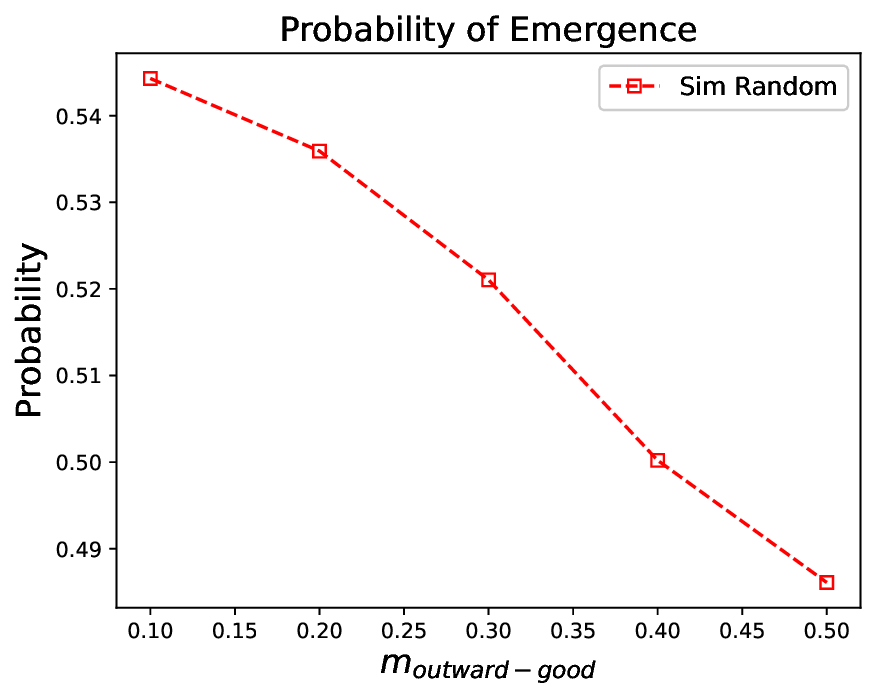}}
    \subfigure[Epidemic Size]{
    \label{fig:es_ccm}
    \includegraphics[width=0.4\textwidth]{ 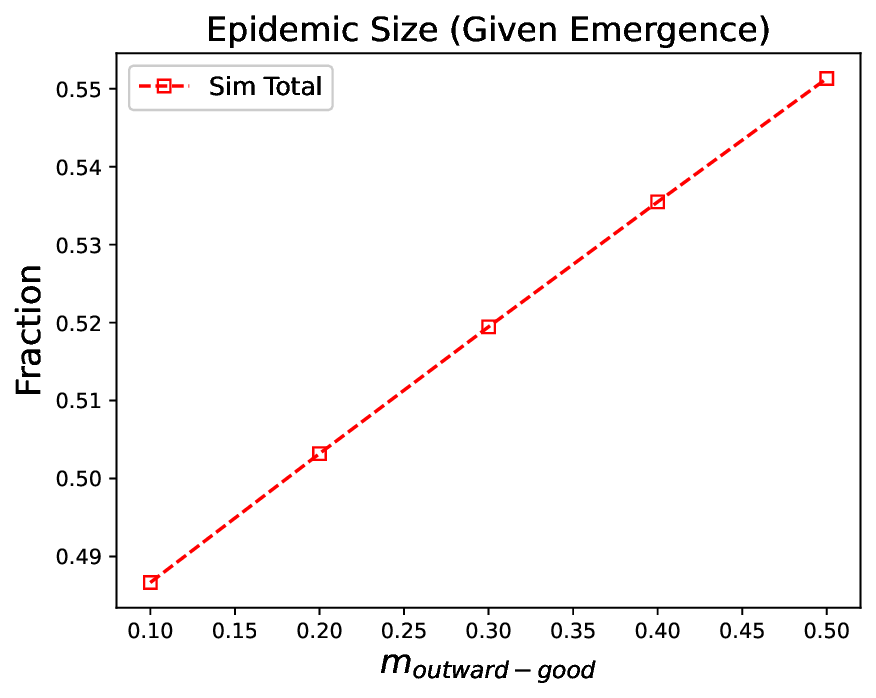}}
    \caption{\sl 
Probability of Emergence (a), Epidemic Size (given emergence) (b) with an increasing proportion of outward-good mask wearers ($m_{outward-good}$) in the population under a mean degree = 5 ($s = 1$ and $t = 2$) using networks generated by configuration model with clustering. The percentage of the no-mask population $x = 40$. 
    $\boldsymbol{\epsilon}_{out} = [0.7, 0.3, 0]$, and $\boldsymbol{\epsilon}_{in}=[0.3, 0.7, 0]$ for inward-good, outward-good and no mask. 
    $T$ = 0.6.
    The simulation is done with $1,000,000 $ nodes and $5,000$ experiments. The average clustering coefficient of the networks is 0.247 (compared to 0.004 for figure \ref{fig:vary_inout_md10_es_pe}).
    The trend that as $m_{outward-good}$ increases, the probability of emergence with random seed decreases and total epidemic size increases still holds even with a higher clustering coefficient. 
    }
    \label{fig:ccm}
\end{figure*}

Figure \ref{fig:ccm} shows the results for experiment setting 3. In this experiment, instead of using the original configuration model to generate the random graphs, we use the generalized \textit{configuration model with clustering} \cite{newman_random_2009, miller_percolation_2009}. 
In this generalized model, for a node $i$, it is required to specify separately $s_i$, the number of single edges (``stubs'') and $t_i$, the number of complete triangles (``corners of triangles'') attached to it.
The joint degree sequence $\{ s_i, t_i\}$ describes the numbers of such stubs and corners for every vertex. 
Given the degree sequence, we can generate the network by selecting pairs of stubs uniformly at random and choosing trios of corners at random to form complete triangles. 
Similar to the constraint on the degree sequence for the original configuration model (the degree sequence has to have an even sum), the sum of the edge degree sequence $\{s_i\}$ has to be even, and the sum of the triangle degree sequence $\{t_i\}$  has to be divisible by 3 for the generalized configuration model.
Given the joint degree sequence $\{ s_i, t_i\}$, let's say on average, each vertex is connected to $s$ single edges and $t$ triangles, the mean degree of a vertex is $k = s + 2t$.
In this experiment, we generate $\{s_i\}$ based on Poisson distribution with a mean of 1, and $\{t_i\}$  based on Poisson distribution with a mean of 2. The resulting network thus has a mean degree of 5. We generated the network with 1,000,000 nodes 5,000 times, with an average clustering coefficient of 0.247. Note that in figure \ref{fig:vary_inout_md10_es_pe}, the average clustering coefficient is 0.004 for 1,000,000 nodes over 5,000 trials.
In figure \ref{fig:pe_ccm} and figure \ref{fig:es_ccm}, we still observe the opposite trend for probability of emergence and epidemic size as the percentage of outward-good masks increases.

\begin{figure*}[!h]
    \centering
    \subfigure[Probability of Emergence]{
    \label{fig:pe_haslemere}
    \includegraphics[width=0.4\textwidth]{ 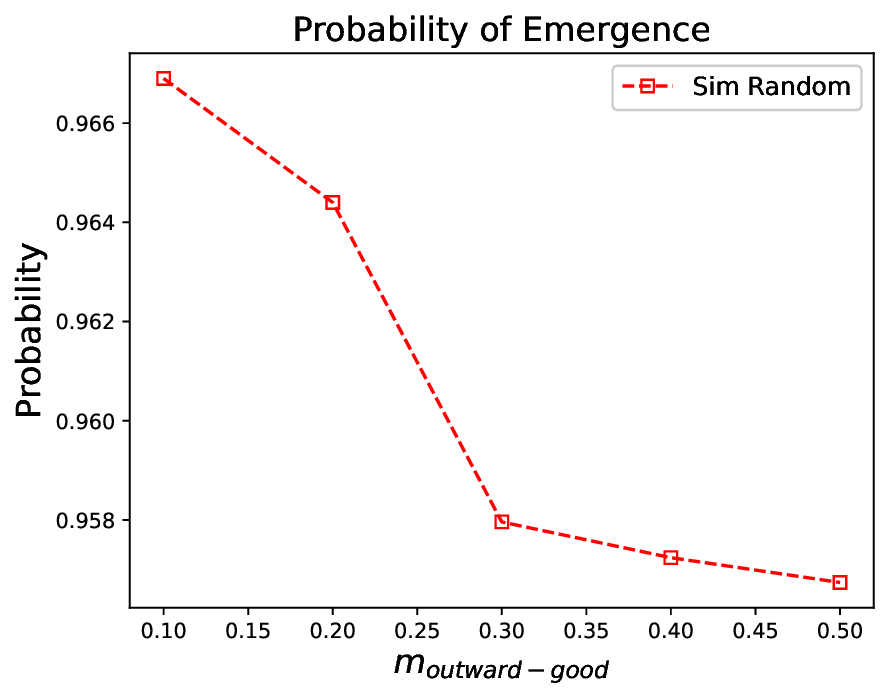}}
    \subfigure[Epidemic Size]{
    \label{fig:es_haslemere}
    \includegraphics[width=0.4\textwidth]{ 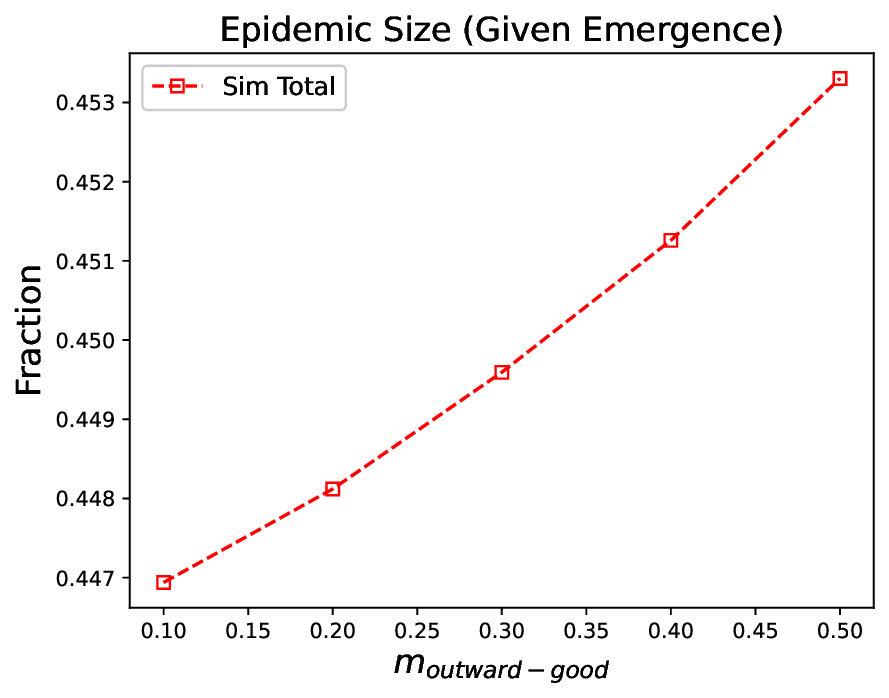}}
    \caption{\sl 
Probability of Emergence (a), Epidemic Size (given emergence) (b) with an increasing proportion of outward-good mask wearers ($m_{outward-good}$) using Haslemere dataset. The percentage of the no-mask population $x = 40$. 
    $\boldsymbol{\epsilon}_{out} = [0.7, 0.3, 0]$, and $\boldsymbol{\epsilon}_{in}=[0.3, 0.7, 0]$ for inward-good, outward-good and no mask. 
    $T$ = 0.6.
    There are 469 nodes and 8277 edges in the network, with a mean degree of 35.29. 
    The simulation is done with $5,000$ experiments. The clustering coefficient of the Haslemere contact network is 0.248 (compared to 0.004 for figure \ref{fig:vary_inout_md10_es_pe}).
    The trend that as $m_{outward-good}$ increases, the probability of emergence with random seed decreases and total epidemic size increases still holds with a higher clustering coefficient on a real-world contact network. }
    \label{fig:haslemere}
\end{figure*}

Figure \ref{fig:haslemere} presents the simulation results on the real-world dataset described by experiment setting 4 \cite{noauthor_sparking_nodate}. The original dataset is composed of pairwise distances between users in the BBC Pandemic Haslemere app over time in the town of Haslemere in UK. The network is undirected, where the edge represents an encounter within 50 mins between two users. For simulating disease spread using \textit{Mask model}, we binarize the edges to 1 and 0. The processed network consists of 469 nodes and 8277 edges, with a mean degree of 35.29, and a clustering coefficient of 0.248. The clustering coefficient here is close to that in experiment setting 3. 
We adopt the same parameter choice as experiment setting 3 except for the network structure.
In figure \ref{fig:pe_haslemere} and figure \ref{fig:es_haslemere}, the probability of emergence decreases as the percentage of outward-good masks increases, and in the meanwhile, epidemic size increases. This shows the similar trends we observe in experiment setting 1 to 3.
Compared to figure \ref{fig:pe_ccm}, the values for probability of emergence in figure \ref{fig:pe_haslemere} are much higher than that in figure \ref{fig:pe_ccm}. But epidemic size for these two experiment settings are not too different. This could be due to a much larger mean degree and a small number of nodes Haslemere dataset has.
\nocite{*}

\end{document}